\documentclass[%
reprint,
 amsmath,amssymb,
 aps,
showkeys,
twocolumn
]{revtex4}%{revtex4-2}

\usepackage{graphicx}% Include figure files
\usepackage{dcolumn}% Align table columns on decimal point
\usepackage{bm}% bold math
\usepackage{hyperref}% add hypertext capabilities
\usepackage{algorithm}
\usepackage{algpseudocode}
\usepackage{algorithmicx}
\usepackage{hhline}

\usepackage{ulem}

\usepackage{xcolor}
\usepackage{soul}

\usepackage{import}
\usepackage{xifthen}
\usepackage{pdfpages}
\usepackage{transparent}

\newcommand{\added}[1]{{{{#1}}}}
\newcommand{\addedaccepted}[1]{{{#1}}}

\begin{document}

\title{{\tt Elliptica}: a new pseudo-spectral code for the construction of initial data}

\author{Alireza Rashti$^1$, Francesco Maria Fabbri$^2$, Bernd Br\"ugmann$^2$,
Swami Vivekanandji Chaurasia$^3$, Tim Dietrich$^{4,5}$,
Maximiliano Ujevic$^6$, Wolfgang Tichy$^1$}
\affiliation{
${}^1$Department of Physics, Florida Atlantic University,
Boca Raton, FL 33431, USA \\
${}^2$Theoretical Physics Institute, University of Jena, 07743 Jena, Germany\\
${}^3$The Oskar Klein Centre, Department of Astronomy, Stockholm University, AlbaNova, SE-10691 Stockholm, Sweden \\
${}^4$Institut f\"{u}r Physik und Astronomie, Universit\"{a}t Potsdam, Haus 28, Karl-Liebknecht-Strasse 24/25, 14476, Potsdam, Germany\\
${}^5$Max Planck Institute for Gravitational Physics (Albert Einstein Institute),
Am M\"{u}hlenberg 1, Potsdam 14476, Germany\\
${}^6$Centro de Ci\^encias Naturais e Humanas, Universidade Federal do ABC, 09210-170, Santo Andr\'e, S\~ao Paulo, Brazil}

\date{\today}

\begin{abstract}
Numerical studies of the dynamics of gravitational systems, e.g.,
black hole-neutron star systems,
require physical and constraint-satisfying initial data.
In this article, we present the newly developed pseudo-spectral code {\tt Elliptica},
an infrastructure for construction of initial data
for various binary and single gravitational systems of all kinds.
The elliptic equations under consideration are solved
on a single spatial hypersurface of the spacetime manifold.
Using coordinate maps, the hypersurface is covered by patches whose
boundaries can adapt to the surface of the compact objects.
To solve elliptic equations with arbitrary boundary condition,
{\tt Elliptica} deploys a Schur complement domain decomposition method
with a direct solver.
In this version, we use cubed sphere coordinate maps and
the fields are expanded using Chebyshev polynomials of the first kind.
Here, we explain the building blocks of {\tt Elliptica} and the
initial data construction algorithm for a
black hole-neutron star binary system. 
We perform convergence tests and evolve the data to validate our results.
Within our framework, the neutron star can reach \addedaccepted{spin values close to breakup
} with arbitrary direction, 
while the black hole can have arbitrary spin with dimensionless spin magnitude
 $\sim 0.8$.
\end{abstract}

\keywords{black hole-neutron star binary, initial data}

\maketitle

%\tableofcontents
  
%┌──────────────┐
%│ introduction │
%└──────────────┘
\section{Introduction}

Observations of gravitational 
waves~\cite{Abbott:2016blz, TheLIGOScientific:2017qsa, %
LIGOScientific:2021qlt, LIGOScientific:2018mvr, Abbott:2020niy}
are treasure troves of information about a broad spectrum of  
large scale physics, such as the nature of gravity%
~\cite{Ezquiaga:2017ekz, Baker:2017hug, Creminelli:2017sry, %
LIGOScientific:2020tif, LIGOScientific:2019fpa, LIGOScientific:2018dkp}%
, and small scale physics, such as properties of the equation of state of supranuclear-dense
matter~\cite{TheLIGOScientific:2017qsa, Abbott:2018wiz, Abbott:2018exr, %
LIGOScientific:2018mvr, De:2018uhw}.
New gravitational wave observing runs of 
the LIGO-Virgo Collaboration and improvements in detectors'
sensitivity~\cite{KAGRA:2013rdx} 
increase the demand for a better understanding of the complex physics
present around the moment of merger of the compact objects. This, on the
other hand, requires high accuracy numerical-relativity simulations.

Any kind of numerical-relativity simulation requires as a starting point
constraint-satisfying and self-consistent initial data~(ID).
In this sense, the accuracy and reliability of the simulations depend 
on the accuracy of the ID.
\addedaccepted{ Because of this, significant efforts have been put
by the entire numerical-relativity community into
developing ID solvers. Among these solvers are: 
the {\tt COCAL} code~\cite{Uryu2012, Tsokaros:2015fea} for constructing 
quasi-equilibrium spinning binary black hole~(BBH) and 
binary neutron star~(BNS) initial data on circular,
but not eccentricity-reduced orbits;
the publicly available {\tt FUKA} code~\cite{Papenfort:2021hod} for computing 
eccentricity-reduced BBH, BNS, and black hole-neutron star~(BHNS) binaries 
with aligned or anti-aligned spins;
the public code {\tt LORENE}~\cite{Lorene, Grandclement2006, Taniguchi:2006yt,
Taniguchi:2007xm, Taniguchi:2007aq} for quasi equilibrium or corotational 
BBH, BHNS, and BNS, where only black holes can have spins that are aligned
or anti-aligned, and of low magnitude;
a private version of {\tt LORENE}~\cite{Kyutoku_2009, Kyutoku_2011, Kyutoku_2021}
for the construction of quasi-equilibrium, eccentricity reduced BBH, BNS, 
and BHNS binaries (but with limited magnitude and direction of spin);
the {\tt NRPyElliptic} code~\cite{Assumpcao:2021fhq} for BBHs;
the {\tt SGRID}~\cite{Tichy_2009, Tichy_2012, Tichy_2019} code,  
capable of producing BNS configurations with arbitrary spin orientation, eccentricity, and mass ratio;
{\tt SpECTRE}'s elliptic solver \cite{Fischer:2021voj,Fischer:2021qbh} for computing BBHs ID;
the private code {\tt Spells}~\cite{Pfeiffer_2003_1, Foucart_2008, Tacik2015, Tacik2016}
for the construction of eccentricity-reduced BBH, BNS, and BHNS binaries
with arbitrary spin and asymmetric masses;
and {\tt TwoPunctures}~\cite{Ansorg2004, Ansorg2005, Khamesra:2021duu}
for BBHs and non-spinning BHNSs.
As the current public ID solvers are limited to aligned or anti-aligned
spin directions for BHNS systems, developing an infrastructure for
the construction of ID with highly spinning black holes or neutron stars
with arbitrary spin directions is important.
}

\addedaccepted{We have developed} a new pseudo-spectral code, {\tt Elliptica}, 
as an infrastructure for the construction of 
ID of various astrophysical compact objects. 
\addedaccepted{{\tt Elliptica}'s framework is such that each compact object, like
a neutron star~(NS) or black hole~(BH), is implemented in a separate module. 
Hence, in principle, one can
combine different modules to create ID for BBH, BNS, and BHNS
systems.}
In this work, we show its usage for the computation of BHNS ID.
{\tt Elliptica}'s general benefits over other available codes are its
suitability for efficient eccentricity reduction, 
\addedaccepted{cf.~\cite{Kyutoku_2021}}, and
its ability to
compute systems in which the NS and BH spins can
point in arbitrary directions, and its ability to compute highly spinning
BHs and NSs.

The organization of the paper is as follows:
in section~\ref{sec_basics} we present the foundation,
such as the coordinate setup and the method used for the solution of
elliptic equations;
in section~\ref{sec_formalism} we cover the formalism employed to
derive the Einstein's constraint equations and Euler's equations
for BHNS systems; 
in section~\ref{sec_numerical_method} we explain the algorithms applied
to construct physical and constraint-satisfying ID for BHNSs;
in section~\ref{sec_results} we validate the code by performing 
convergence tests of BHNS ID,
\addedaccepted{by comparing with analytical 
approximations, and by performing dynamical evolution} of the data; 
in section~\ref{sec_summary} we present our conclusions.
We use geometric units with $G = c = M_{\odot} = 1$ in this paper.
Summation over repeated indices is implied unless otherwise mentioned.

%┌─────────────────────────────────────┐
%│ numerical details of Elliptica code │
%└─────────────────────────────────────┘
\section{{\tt Elliptica}'s foundation}
\label{sec_basics}

\subsection{Overview}

{\tt Elliptica} is designed to construct ID for equilibrium and 
quasi-equilibrium astrophysical systems composed of single or 
binary compact objects which can be BHs or NSs.
To construct the ID, one has to solve
the constraint equations together with Euler's equations.
The constraint equations are derived from Einstein's equations,
and Euler's equations are derived from the conservation of the
stress-energy tensor and the continuity equation; cf.~\ref{sec_formalism}.
\addedaccepted{These equations can be cast into the form of
hyperbolic-parabolic or hyperbolic-algebraic system 
such as~\cite{Racz:2015ena,Racz:2015mfa,Racz:2017krc,Csukas:2019qco}.
They can also be put into hyperbolic form
and evolved forward in an unphysical time to find a steady state solution that
satisfies the equations~\cite{Ruter:2017iph,Assumpcao:2021fhq}.
However, in this work we express these constraint equations in
the form of coupled nonlinear elliptic partial differential equations~(PDE)s.}
Additionally, the NS surface location is not known a priori.
This means these coupled elliptic PDEs need to be augmented by an algebraic
equation for the NS surface. 
Therefore,\addedaccepted{ in this work,} making ID is tantamount to a procedure to find 
the solution of these equations (elliptic plus algebraic equations).
Furthermore, since the ID 
are sought for a specific physical system with specific properties, 
the solution must be guided towards these physical parameters
throughout this procedure; cf.~\ref{sec_numerical_method}.

{\tt Elliptica} has been written completely in the {\tt C} programming
language, but an extensive use of structures has allowed for the
incorporation of some object-oriented design principles.
It currently supports only shared-memory multiprocessing. 
To generate numerical-relativity equations in {\tt C},
the open source code {\tt Cpi} has been used~\cite{alireza_rashti_2021_5495885}
(but other means can be used too).

As an infrastructure for construction of ID, 
{\tt Elliptica} requires two main components, an elliptic solver and a
computational grid.
In this section we present these ingredients by explaining 
how an elliptic equation is set up and solved using a pseudo-spectral
method~\cite{boyd1989chebyshev, Grandclement2009}
together with a Schur complement domain decomposition~(SCDD) method%
~\cite{saad2003iterative}. Moreover, we \added{derive analytical expressions for a fast
computation of} 
the Jacobian of an elliptic equation. Finally, we illustrate our computational grid.

\subsection{Elliptic Solver}

A key ingredient of \addedaccepted{many} ID codes is a 
routine that solves elliptic equations with given boundary conditions.
However, this undertaking is often computationally expensive. 
To reduce the overhead, we \added{take advantage of spectral techniques} to efficiently
compute the Jacobian needed when linearizing elliptic equations.

\subsubsection{Jacobian Matrix}

To explain the idea, let us consider solving the $1$-dimensional 
Poisson equation
\begin{equation}
        \nabla^2 u = 
        \frac{\partial^2 u}{\partial x^2} = S,
        \, \, u\vert_{\partial \Omega} = 0.
\end{equation}
for the field $u=u(x)$ with source $S=S(x)$
on a computational grid $\Omega=\{x:x\in[a,b]\}$. Here $a$ and $b$ are
real numbers and $\partial \Omega$ denotes the boundary.
First, we discretize the problem by introducing $N$ grid points.
Now, instead of one PDE we thus obtain one algebraic equation per 
grid point:
\begin{equation}
        F_i(\vec{u}) := \{\nabla^2 u - S\} {\vert_i} = 0,
	\label{eq_Fi}
\end{equation}
here, $i$ refers to the index of a grid point, 
$\vec{u} = (u_0,u_1,...,u_{N-1})^T$ is $u$ at each grid point,
and $\nabla^2 u {\vert_i}$ is the discretization of the derivative
$\nabla^2 u$ at the grid point $i$.
The Newton-Raphson method also requires the linearization of this
equation. This involves the computation of the Jacobian matrix
\begin{equation}
 J_{ij} := \frac{\delta F_i(\vec u)}{\delta u_j},
 \label{eq_def_jacobian_matrix}
\end{equation}
where indices $i$ and $j$ refer to the indices of the grid points
after the discretization. A Newton-Raphson step then consists of
$u \rightarrow u+\tilde{u}$, 
where $\tilde{u}$ is the solution of the linear equation $J \tilde{u} = -F$.
The method starts from an initial guess for $u$ and iterates until a given
stopping criterion is met, e.g., until a desired tolerance for the $L^2$
norm $||\vec{F}||$ is reached or if the number of iterations exceeds some
limit. The main steps of our Newton-Raphson 
procedure~\cite{burden2011numerical,numerical_recipes} 
is summarized in  Algorithm~\ref{algo_newton}.

%┌───────────────┐
%│ Newton method │
%└───────────────┘
\begin{figure}[t]
\begin{algorithm}[H]
\caption{Newton-Raphson algorithm:
         the number of iterations $N_{iter}$ and the desired tolerance $tol$ 
         are given.}
\label{algo_newton}
\begin{algorithmic}[1]
  \State set $k=0$
  \While{ $(k \leq N_{iter}$ or 
          $\;||\vec F(\vec u)|| \geq tol)$\;}
  \State compute $F_i(\vec u) = \{\nabla^2 u-S\} \vert_{i}$
         at each grid point $i$;
  \State compute Jacobian matrix 
         $J_{ij} = \frac{\delta F_i(\vec u)}{\delta u_j}$;
  \State solve matrix equation $J \tilde{u} = -F$ for $\tilde{u}$;
  \State set $u_i \rightarrow u_i+\tilde{u}_i$;
  \State set $k \rightarrow k + 1$;
  \EndWhile
\end{algorithmic}
\end{algorithm}
\end{figure}

In many cases constructing the matrix $J_{ij}$
(defined in Eq.~(\ref{eq_def_jacobian_matrix})) from an analytical 
expression is not practical. 
Instead one often uses a finite difference approximation:
\begin{equation}
         \label{jacobian_def}
        J_{ij}  = \frac{\delta F_i(\vec u)}{\delta u_j}
        \approx \frac{F_i(\vec u + h\vec e_j)-F_i(\vec u)}{h},
\end{equation}
where $h$ is a small value, on the order of the grid spacing, and  
$\vec e_j$ is the vector whose only nonzero component equals to $1$ 
in its j-th entry.
For a 1-dimensional problem, the time complexity of this method is
about $O(N\times N\ln N)$ in which $N$ is the number of grid points and
 $N\ln N$ comes from a fast Fourier transformation needed to compute
the derivatives for each change $\vec u \rightarrow \vec u+h\vec e_j$.
However, one can calculate this expression not only exactly in 
a closed form but also faster with time complexity of order $O(N^2)$ 
by using a spectral expansion.
In the following we explain this more efficient method for 
calculating $J_{ij}$.

\subsubsection{Spectral Jacobian}

{\tt Elliptica} currently uses Chebyshev polynomials of the first kind 
$T_i(X) = \cos(i \arccos(X))$ as the basis of the 
spectral expansion~\cite{boyd1989chebyshev, Grandclement2009},
where $X = \frac{2x -(a+b)}{a-b}$ 
and $i=0,1,\hdots,N-1$.
As the collocation points, it uses the extrema of the 
Chebyshev polynomial $T_{N-1}(X)$ which are
$X_i = \cos\left(\frac{i\pi}{N-1}\right)$ for each $i$.
As discussed in subsection~\ref{sec_grid}, a further
coordinate transformation can be used to map the $x$ coordinates to
other coordinates that are better adapted to the domain shapes 
we intend to use. Here, for the sake of simplicity, we ignore
this transformation but the generalization is straightforward.
The field $u$ as a function of $X$ can then be approximated as
\begin{align}
        u(X) &= \sum_{n=0}^{N-1}{\eta_n c_n T_n(X)},
\end{align}
and consequently, the value $u_i = u(X_i)$ at each grid point $i$ is given by
\begin{align}
        u_i &= \sum_{n=0}^{N-1}{\eta_n c_n T_n(X_i)},
\end{align}
where,
\begin{align}
        \eta_n & = 
        \begin{cases}
                1,\ \ \ \text{if}\; n=0\; \text{or}\; n=N-1
                \\
                2,\ \ \ \text{otherwise}
        \end{cases},
        \nonumber \\
        c_n &= \frac{1}{2(N-1)}
        \sum_{k=0}^{N-1}{\eta_k  u_k  T_k(X_n)}.
        \label{cheb_expansion}
\end{align}
Moreover, the second order derivative of $u(x)$ with respect to $x$ and 
the variation of $c_n$ with respect to $u_j = u(x(j))$, 
which are needed to calculate $J_{ij}$, read
\begin{align}
    \frac{d^2 u_i}{dx^2} &= 
    \left(\frac{dX}{dx}\right)^2 \sum_{n=0}^{N-1} \eta_n  
    c_n  \frac{d^2}{dX^2} T_n(X)|_{X=X_i},
    \label{eq_second_der}
    \\
    \frac{\delta c_n}{\delta u_j} &=\frac{1}{2(N-1)}
    \sum_{k=0}^{N-1}{\eta_k  \delta_{jk} T_k(X_n)}
    \nonumber \\
    &=\frac{\eta_j  T_j(X_n)}{2(N-1)}.
    \label{eq_coeff_var}
\end{align}
Consequently, $J_{ij}$ is:
\begin{align}
        J_{ij}&= \frac{\delta F_i(\vec u)}{\delta u_j}
        \nonumber \\
        &= \frac{\delta}{\delta u_j} \frac{d^2 u_i}{dx^2}
          +
          \overbrace{\frac{\delta}{\delta u_j} S(x)}^{=0}
        \nonumber \\
        &= \frac{\eta_j }{2(N-1)}\left(\frac{dX}{dx}\right)^2 
	\times 
	\nonumber \\
	& \left\{\sum_{n=0}^{N-1} \eta_n
        \ T_j(X_n)  \frac{d^2}{dX^2}  T_n(X)|_{X=X_i}
	\right\},
        \label{variation}
\end{align}
where in the third line we have used Eq.~(\ref{eq_coeff_var}).
Note that the sum in Eq.~(\ref{variation}) can be written as
\begin{align}
        &\sum_{n=0}^{N-1} \eta_n
        \ T_j(X_n)  \frac{d^2}{dX^2}  T_n(X)|_{X=X_i} = &
	\nonumber \\
	&\frac{\partial^2}{\partial X^2}	
	\bigl\{
	2\sum_{n=0}^{N-1} T_n(X_j) T_n(X) 
        \nonumber \\
	&- T_j(X_0)T_0(X) - T_j(X_{N-1})T_{N-1}(X)
	\bigr\} |_{X=X_i}
        \label{eq_j_sum}
\end{align}
where we have used the definition of $\eta_n$ in Eq.~(\ref{cheb_expansion})
and the relation $T_n(X_i) = T_i(X_n)$ which holds since 
\begin{equation}
	T_i(X_n) = 
        \cos\left(i\arccos(X_n)\right) = 
        \cos \left(\pi \frac{ i \, n}{N-1}\right).
	\label{eq_t_symm}
\end{equation}
Additionally, we have changed the notation in
Eq.~(\ref{eq_j_sum}) to emphasize that the derivative only acts on $T_n(X)$
which is then evaluated at $X=X_i$ as in Eq.~(\ref{variation}).
To further simplify the summation in Eq.~(\ref{eq_j_sum}) we use
\begin{align}
	&2\sum_{n=0}^{N-1} T_n(X_j) T_n(X_i) =
	2\sum_{n=0}^{N-1} \cos(n\theta_j)\cos(n\theta_i) = 
	\nonumber %\label{eq_cos_mul}
	\\ 
	&\sum_{n=0}^{N-1} \cos(n(\theta_i+\theta_j)) +
        \sum_{n=0}^{N-1} \cos(n(\theta_i-\theta_j)) .
	\label{eq_cos_sum}
\end{align}
Here we have defined $\theta = \arccos(X)$ so that
$\theta_i = \frac{\pi i}{N-1}$ and 
$\theta_j = \frac{\pi j}{N-1}$.
Using the following identity~\cite{math_table2014}:
\begin{equation}
    \label{cos_identity}
    \sum_{n=0}^{N} \cos(n\theta) = 
    \frac{1}{2}+
    \frac{\sin((N+\frac{1}{2})\theta)}{2\sin(\frac{\theta}{2})},
\end{equation}
we write $J_{ij}$ in a closed form (note $X_i = \cos(\theta_i)$):
\begin{align}
        \label{Jacobian}
        J_{ij} = & \frac{\eta_j}{2(N-1)} \left(\frac{dX}{dx}\right)^2
        \times \nonumber         \\  \Bigl\{
                 & \frac{\partial^2}{\partial X_i^2}
        \left(
        \frac{\sin((N-\frac{1}{2})(\theta_i+\theta_j))}
        {2\sin(\frac{\theta_i+\theta_j}{2})}
        \right)
        + 
        \nonumber \\
                 &
                 \frac{\partial^2}{\partial X_i^2}
        \left(
        \frac{\sin((N-\frac{1}{2})(\theta_i-\theta_j))}
        {2\sin(\frac{\theta_i-\theta_j}{2})} 
        \right) -
	\nonumber \\
	&
        (-1)^{j}\frac{d^2}{dX^2} T_{N-1}(X)|_{X=X_i}\Bigr\}.
\end{align}
Note that the derivatives $\frac{\partial}{\partial X_i}$ only act on
$\theta_i$.

Some remarks are in order.
First, the generalization to higher dimensions is straightforward,
and analogous to a Chebyshev expansion in higher dimensions.
\addedaccepted{Second, we note that there is no singularity in Eq.~(\ref{eq_j_sum}).
Hence, for cases such as $\theta_i+\theta_j = 0$ or $\theta_i-\theta_j = 0$ 
when Eq.~(\ref{Jacobian}) becomes singular, we use Eq.~(\ref{eq_cos_sum})
to compute $J_{ij}$ (i.e., we do not use the identity in
Eq.~(\ref{cos_identity})).}
\addedaccepted{Third, during the construction of ID, resolution is only
gradually increased, so that many iterations of the elliptic solver per
resolution are required.}
Since Eq.~(\ref{Jacobian}) only depends on the number of grid points,
the piece of $J_{ij}$ coming from derivative operators, 
here $\nabla^2$, remains unchanged at each resolution.
Thus, $J_{ij}$ is calculated only once and used without any 
changes in an iterative scheme.
\addedaccepted{Furthermore,} we note that
the functional derivatives do not act on 
the Jacobian of a coordinate transformation.
Thus, if there are more (and possibly nonlinear)
coordinate transformations then the computation of $J_{ij}$ 
involves similar steps except that some extra terms and coefficients
(coming from the coordinate transformation) need to be included.
\addedaccepted{%
Lastly, let us illustrate how the Jacobian of a more complicated 
equation is calculated. For instance, we assume the Jacobian of the
following equation is needed:
\begin{equation}
  f_2(u)\frac{d^2 u(x)}{dx^2} + 
  f_1(u)\frac{d u(x)}{dx} + f_0(u) = S(x),
\end{equation}
where $f_0(u)$, $f_1(u)$, and $f_2(u)$ 
are (possibly non-linear) functions of $u$.
Hence, similar to Eq.~(\ref{eq_Fi})
\begin{equation}
  F_i(\vec{u}) = \{f_2(u)\frac{d^2 u(x)}{dx^2} + 
  f_1(u)\frac{d u(x)}{dx} + f_0(u) - S(x)\} {\vert_i} = 0.
\end{equation}
Consequently, the Jacobian reads 
(no implied summation on the repeated indices)
\begin{align}
J_{ij} = & f_2(u_i) \frac{\delta}{\delta u_j} \frac{d^2 u_i}{dx^2} +
         \delta_{ij} \frac{d f_2(u_i)}{du} \frac{d^2 u_i}{dx^2} +
\nonumber \\
         & f_1(u_i) \frac{\delta}{\delta u_j} \frac{d u_i}{dx}  +
         \delta_{ij} \frac{d f_1(u_i)}{du} \frac{d u_i}{dx} +
         \delta_{ij} \frac{d f_0(u_i)}{du},
\end{align}
in which, $\delta_{ij}$ is the Kronecker delta and
$\frac{\delta}{\delta u_j} \frac{d^2 u_i}{dx^2}$ term is 
calculated by Eq.~(\ref{Jacobian}).
The terms $\delta_{ij} \frac{d f_0(u_i)}{du}$, 
$\delta_{ij} \frac{d f_1(u_i)}{du}$, and 
$\delta_{ij} \frac{d f_2(u_i)}{du}$ are the analytic functional 
derivative of $f_0(u_i)$, $f_1(u_i)$, and $f_2(u_i)$ respectively.
To compute $\frac{\delta}{\delta u_j} \frac{d u_i}{dx}$ we follow 
the same steps as involved in the calculation of 
$\frac{\delta}{\delta u_j} \frac{d^2 u_i}{dx^2}$ but instead of 
second order derivatives with respect to $x$ we have first order derivatives.
}

In conclusion, we have presented 
a fast and analytic method to compute the Jacobian of an 
elliptic PDE using a spectral method.
\added{While this spectral Jacobian method is relatively straightforward, 
we are not aware of any prior publication about it.}
In the next subsection we discuss how the system
$J \tilde{u} = -F$ is solved.

\subsection{Matrix Solver}
\label{sec_matrix_solver}

Having found $F_i(\vec u)$ and $J_{ij}$, the Newton-Raphson
algorithm~\ref{algo_newton} requires us to solve the matrix equation 
$J \tilde{u} = -F$. The size of the matrix $J$ depends on the
resolution, which is chosen to fit the problem under study and the
coordinate patches being used. For a real production run, 
the size could be as high as $10^{5} \times 10^{5}$.
Direct solvers~\cite{davis2006direct} are generally inefficient for
matrix equations with such large dimensions, thus
iterative solvers~\cite{saad2003iterative} with proper preconditioners 
are commonly used in these cases.
However, direct solvers tend to be more robust and 
predictable and do not require 
preconditioners as opposed to iterative solvers.
As a result, direct solvers are preferred when feasible.
A possible strategy is to divide this big system of equations into
mutually exclusive and collectively exhaustive subsystems 
with smaller dimensions, then instead of solving the whole system at once,
one can separately solve these small subsystems which would enable 
the usage of direct solvers. 
To reach this goal, {\tt Elliptica} employs the
SCDD method, explained 
in~\cite{saad2003iterative}, 
to efficiently solve $J \tilde{u} = -F$ by a direct solver.

\subsubsection{Domain Decomposition Method}

In general, domain decomposition methods, 
and in particular the SCDD method, use a
divide-and-conquer principle to reduce the dimension of a matrix equation.
Thus, one can use direct solvers in a parallel fashion 
to solve the whole matrix equation at once,
which otherwise would have been infeasible due to the very large 
dimension. In this section we demonstrate the gist of the
SCDD method used in {\tt Elliptica}.

Often, a given manifold, here the computational grid, cannot be covered
by a single patch. 
A known example is a $2$-sphere and its 
pole singularities~\cite{frankel2012the}.
\addedaccepted{Moreover, for a spectral method, one desires
to separate matter and vacuum regions into different patches
to avoid Gibbs phenomena~\cite{boyd1989chebyshev}.}
For instance, the NS and the outside of the NS should 
be covered by different patches. Furthermore, it is generally 
required to use different resolutions for different parts of the grid
or to compactify the outer-boundary of the computational grid to
possibly cover spatial infinity. Therefore, different patches with 
different properties are needed and it is natural to cover a grid with 
multiple patches. We exploit this property by using the SCDD method
to solve equations on each of these patches separately.

\begin{figure}[t]
    \centering
    \def\svgwidth{\columnwidth}
%% Accompanies image file 'schur_grid.pdf' (pdf, eps, ps)
\begingroup%
  \makeatletter%
  \providecommand\color[2][]{%
    \errmessage{(Inkscape) Color is used for the text in Inkscape, but the package 'color.sty' is not loaded}%
    \renewcommand\color[2][]{}%
  }%
  \providecommand\transparent[1]{%
    \errmessage{(Inkscape) Transparency is used (non-zero) for the text in Inkscape, but the package 'transparent.sty' is not loaded}%
    \renewcommand\transparent[1]{}%
  }%
  \providecommand\rotatebox[2]{#2}%
  \newcommand*\fsize{\dimexpr\f@size pt\relax}%
  \newcommand*\lineheight[1]{\fontsize{\fsize}{#1\fsize}\selectfont}%
  \ifx\svgwidth\undefined%
    \setlength{\unitlength}{805.865251bp}%
    \ifx\svgscale\undefined%
      \relax%
    \else%
      \setlength{\unitlength}{\unitlength * \real{\svgscale}}%
    \fi%
  \else%
    \setlength{\unitlength}{\svgwidth}%
  \fi%
  \global\let\svgwidth\undefined%
  \global\let\svgscale\undefined%
  \makeatother%
  \begin{picture}(1,0.22793)%
    \lineheight{1}%
    \setlength\tabcolsep{0pt}%
    \put(0,0){\includegraphics[width=\unitlength,page=1]{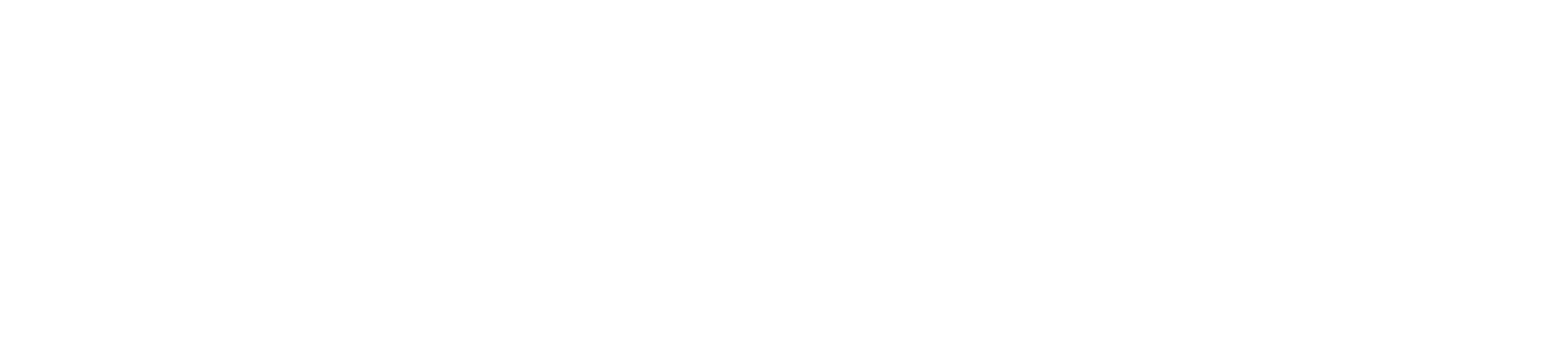}}%
    \put(0.62729081,0.18348984){\color[rgb]{0,0,0}\makebox(0,0)[lt]{\lineheight{1.25}\smash{\begin{tabular}[t]{l}$\Omega_{1}$\end{tabular}}}}%
    \put(0.88121607,0.18348984){\color[rgb]{0,0,0}\makebox(0,0)[lt]{\lineheight{1.25}\smash{\begin{tabular}[t]{l}$\Omega_{2}$\end{tabular}}}}%
    \put(0.01599023,0.00361364){\color[rgb]{0,0,0}\makebox(0,0)[lt]{\lineheight{1.25}\smash{\begin{tabular}[t]{l}$\partial \Omega$\end{tabular}}}}%
    \put(0.1565122,0.12763096){\color[rgb]{0,0,0}\makebox(0,0)[lt]{\lineheight{1.25}\smash{\begin{tabular}[t]{l}$\Omega$\end{tabular}}}}%
    \put(0.71098555,0.07744621){\color[rgb]{0,0,0}\makebox(0,0)[lt]{\lineheight{1.25}\smash{\begin{tabular}[t]{l}$\Gamma$\end{tabular}}}}%
    \put(0,0){\includegraphics[width=\unitlength,page=2]{schur_grid.pdf}}%
    \put(0.60866955,0.00361364){\color[rgb]{0,0,0}\makebox(0,0)[lt]{\lineheight{1.25}\smash{\begin{tabular}[t]{l}$\partial \Omega$\end{tabular}}}}%
  \end{picture}%
\endgroup%

    \caption{%
    An example of a $2$-dimensional grid 
    $\Omega$ \addedaccepted{that} is covered by two subdomains $\Omega_1$ and $\Omega_2$ 
    \addedaccepted{with a} common interface $\Gamma$.
    \addedaccepted{If we solve an elliptic equation on the two subdomains,
    boundary conditions on $\partial \Omega$ and $\Gamma$ are 
    required in order to guarantee a unique solution.
    The boundary conditions in Eqs.~\ref{eq_inner_bc_1} 
    and \ref{eq_inner_bc_2} at $\Gamma$ couple the elliptic equations
    in the subdomains $\Omega_1$ and $\Omega_2$ together.
    Yet, once the values of the fields are known on $\Gamma$,
    the elliptic equations in the two subdomains can be decoupled and solved
    separately.}}
    \label{fig_schur_grid}
\end{figure}

To demonstrate the idea of SCDD, consider solving a
$2$-dimensional elliptic equation on the grid $\Omega$, shown in 
Fig.~\ref{fig_schur_grid}, 
with a some boundary condition on $\partial \Omega$. 
The elliptic equation has a unique solution if appropriate 
boundary conditions~(BCs) are imposed.
Thus, if we attempt to solve this equation 
separately on each subdomain the system would be under-determined 
because the BCs on $\Gamma$ are not known yet.
Therefore we impose the following BCs to close the 
system~\cite{Tichy_2019}:
\begin{align}
        \vec n\cdot \vec \nabla u_{(\Omega_1)}\vert_{\Gamma} -  
        \vec n\cdot \vec \nabla u_{(\Omega_2)}\vert_{\Gamma} & = 0
        \label{eq_inner_bc_1}
    , \\
        u_{(\Omega_1)}\vert_{\Gamma} -   
        u_{(\Omega_2)}\vert_{\Gamma} & = 0,
        \label{eq_inner_bc_2}
\end{align}
where $\vec n$ is the normal vector to the common interface $\Gamma$, and
$u_{(\Omega_1)}\vert_{\Gamma}$ and $u_{(\Omega_2)}\vert_{\Gamma}$
denote the solution from domain $\Omega_1$ and $\Omega_2$
evaluated on the interface $\Gamma$.
%In practice, Eq.~(\ref{eq_inner_bc_1}) is imposed on the boundary grid
%points of domain $\Omega_{1}$ and Eq.~(\ref{eq_inner_bc_2}) is imposed
%on the boundary grid points of domain $\Omega_{2}$. 
Eqs.~(\ref{eq_inner_bc_1}) and (\ref{eq_inner_bc_2})
create a coupling between the interior of each subdomain and 
the interface and vice versa. Hence, if the solution was known on the
interface $\Gamma$, the problem would be reduced
to solve two uncoupled elliptic equations in each subdomain. 
Therefore, to decouple the system it is natural to find the solution on
the interface $\Gamma$ first.
This decoupling is the main idea of the SCDD method.

\subsubsection{Schur Domain Decomposition Method}

We consider a grid $\Omega$ with outer-boundary $\partial \Omega$
which is covered by subdomains (patches)                                 
$\Omega_1,\, \Omega_2,\,\hdots,\, \text{and}\,\Omega_s$, 
where $s$ is the number of subdomains.
Moreover, any two subdomains might have one or more common interfaces
which results in a coupling of the two subdomains.
As we mentioned earlier the goal is to find the solution on the common
interfaces first (to decouple them) and then solve the elliptic equation 
for each patch independently. 
Therefore, following the algorithm~\ref{algo_newton} 
after setting up the matrix equation $J \tilde{u} = -F$,
we reorder this system of equations, such that it has the following
general structure:
\begin{equation}
\label{eq_J_reordered}
\begin{aligned}
\begin{pmatrix}
B_{1}  &        &        &        & E_{1}  &        &        &         \\
       & B_{2}  &        &        &        & E_{2}  &        &         \\
       &        & \ddots &        &        &        & \ddots &         \\
       &        &        & B_{s}  &        &        &        & E_{s}   \\
F_{11} & F_{12} & \hdots & F_{1s} & C_{11} & C_{12} & \hdots & C_{1s}  \\
F_{21} & F_{22} & \hdots & F_{2s} & C_{21} & C_{22} & \hdots & C_{2s}  \\
\vdots & \vdots & \vdots & \vdots & \vdots & \vdots & \vdots & \vdots  \\
F_{s1} & F_{s2} & \hdots & F_{ss} & C_{s1} & C_{s2} & \hdots & C_{ss}  \\
\end{pmatrix}
\begin{pmatrix}
%\overbrace{x_{1}}^{all\,but\,inner\,boundary\,points}   \\
\tilde v_{1}   \\     
\tilde v_{2}   \\
\vdots         \\
\tilde v_{s}   \\
\tilde w_{1}   \\     
\tilde w_{2}   \\
\vdots         \\
\tilde w_{s}   \\
%\underbrace{y}_{inner\,boundary\,points}
\end{pmatrix}
 =
\begin{pmatrix}
f_{1}   \\
f_{2}   \\
\vdots  \\
f_{s}   \\
g_{1}   \\
g_{2}   \\
\vdots  \\
g_{s}   \\
\end{pmatrix},
\end{aligned}
\end{equation}
where empty entries are zero. Let us define $p,q = 1,\hdots,s$. Then,
the $B_p$ are the \addedaccepted{sub-matrices} of the matrix $J$ that are interior 
to the $p$-th subdomain
and include outer boundary conditions~(if any exist for this
subdomain). The $E_p$ represent the \addedaccepted{sub-matrices} of the matrix $J$ due to
the coupling of interior and interface of the $p$-th subdomain. 
The $F_{pq}$ describe the coupling between the $p$-th interface and
the $q$-th interior.
Finally, $C_{pq}$ represent all of the couplings between the 
$p$-th interface to the $q$-th interface.
Furthermore, each $\tilde v_p$ is the subvector of unknowns that is
interior to the $p$-th subdomain and 
each $\tilde w_p$ represents the subvector of unknowns for all interfaces.
Each $f_p$ shows the source term portion in the $p$-th subdomain and
each $g_p$ is the left hand side of Eq.~(\ref{eq_inner_bc_1}) or
Eq.~(\ref{eq_inner_bc_2}) depending which one has been imposed for
the corresponding subdomain.

In this new arrangement of the matrix $J$, all of 
the coupled and unknown values are encapsulated in the subvector
$\tilde w=(\tilde w_1,\tilde w_2,\hdots,\tilde w_s)^T$.
Thus, finding these values results in decoupling of the system.
To better explain the idea, it is convenient to write 
Eq.~(\ref{eq_J_reordered}) like a system of equations with 
two unknowns as follows:
\begin{align}
\label{eq_schur_2d}
%\Rightarrow \text{in effect:}\;
\begin{pmatrix}
B & E \\
F & C
\end{pmatrix}
\begin{pmatrix}
\tilde v \\
\tilde w \\
\end{pmatrix}
& =
\begin{pmatrix}
f \\
g \\
\end{pmatrix}.
\end{align}
In order to find $\tilde w$ we write the first row in Eq.~(\ref{eq_schur_2d}) as
\begin{equation}
\label{eq_x_eq}
B \tilde v + E \tilde w  = f \Rightarrow  \tilde v = B^{-1}(f-E \tilde w).
\end{equation}
Then by substituting $\tilde v$ in the second row of Eq.~(\ref{eq_schur_2d}),  
we find the reduced system
\begin{equation}
\label{eq_y_eq}
F \tilde v+C \tilde w  = 
g \Rightarrow  (C-FB^{-1}E) \tilde w  = 
g -F B^{-1}f.
\end{equation}
The solution $\tilde w$ of Eq.~(\ref{eq_y_eq}) can now
be found independently of the values of $\tilde v$ and then $\tilde w$
can be back substituted into Eq.~(\ref{eq_x_eq}) to find $\tilde v$.

In practice, we implement the following equations for each
subdomain $p$ 
(no summation is implied on the repeated indices)~\cite{saad2003iterative}:

\begin{align}
        B_p \tilde v_p + E_p \tilde w_p &= f_p,
	\label{eq_xi_eq}
        \\
%        \sum_{j=1}^{s}(F_{ij} x_j + C_{ij} y_j) &= g_i.
%	\label{eq_yi_eq}
 \sum_{q=1}^{s}(C_{pq}-F_{pq} E_q^\prime )\tilde w_q &= 
        g_p-\sum_{q=1}^{s}(F_{pq}f_q^\prime),
	\label{eq_syi_eq}
\end{align}
where
\begin{align}
        E_p^\prime &= B_p^{-1} E_p,
        \nonumber\\
        f_p^\prime &= B_p^{-1} f_p.
	\label{eq_Ep_fp}
\end{align}
%
%Eq.~(\ref{eq_yi_eq}) results in:
%
%\begin{align}
%        \sum_{j=1}^{s}(C_{ij}-F_{ij} E_i^\prime )y_j = 
%        g_i-\sum_{j=1}^{s}(F_{ij}f_j^\prime).
%	\label{eq_syi_eq}
%\end{align}
%
The system of equations~(\ref{eq_syi_eq}) only involves $\tilde w$, i.e.,
$\tilde{u}$ at the interface points. It can be summarized as:
\begin{equation}
        \label{eq_schur_y}
        S \tilde w = g^\prime ,
\end{equation}
where, $S$ is called Schur complement matrix.
After solving Eq.~(\ref{eq_schur_y}) 
for $\tilde w_p$, $\tilde v_p$ is found using
Eq.~(\ref{eq_xi_eq}) which can be written as
$\tilde v_p  = f_p^\prime - E_p^\prime \tilde w_p$.
A summary of SCDD method to solve elliptic equations is               
shown in algorithm~\ref{algo_schur}.

\begin{figure}[t]
\begin{algorithm}[H]
\caption{Schur complement domain decomposition method.}
\label{algo_schur}
\begin{algorithmic}[1]
	\State Solve $B E^\prime = E$ for $E^\prime$;
	\State Solve $B f^\prime = f$ for $f^\prime$;
	\State Compute $g^\prime = g-Ff^\prime$;
	\State Compute $S=(C-FE^\prime)$;
	\State Solve $S \tilde w = g^\prime$ for $\tilde w$;
	\State Compute $\tilde v=f^\prime -E^\prime \tilde w$;
\end{algorithmic}
\end{algorithm}
\end{figure}

Regarding the implementation of SCDD in {\tt Elliptica},
a few comments are in order.
First, the system of equations is set up
in the same order as Eq.~(\ref{eq_J_reordered}), thus, no overhead 
for reordering of the system is incurred.
Second, in Eq.~(\ref{eq_Ep_fp}) instead of inverting the matrix $B_p$
directly, equations $B_p E^{\prime}_{p} = E_p$ and $B_p f^{\prime} = f_p$
are solved for $E^{\prime}_{p}$ and $f^{\prime}$, respectively.
Also, we note that these calculations can be
performed in parallel since $B_p$'s describe the uncoupled blocks
in the $J$ matrix.
Third, many columns in the $F_{pq}$ and $C_{pq}$ matrices
are \addedaccepted{equal to} zero 
because each subdomain $p$ only has interfaces where it touches
other subdomains, which only happens for neighboring subdomains. 
Since all the matrices are stored in 
compressed column storage~\cite{davis2006direct},
there is no computational cost for setting zero entries of $F_{pq}$ and 
$C_{pq}$.
Fourth, the Schur complement matrix consists of the coupling 
information coming from the interfaces, thus its dimension is as big as
the total number of points on all interfaces.
If we use the cubed spherical grid of Fig.~\ref{fig_full_cubed_sphere}
with $20$ points in each direction in each patch, the matrix $S$ is
approximately $56000$-dimensional.
Nevertheless it is quite sparse
and hence, the use of an efficient sparse solver is feasible to solve
this matrix equation.
Lastly, to invert $B_p$ matrices and solve Eq.~(\ref{eq_schur_y})
the open source unsymmetric multifrontal direct
solver {\tt UMFPACK}~\cite{umfpack} is used.

\begin{figure}[t]
    \centering
    \def\svgwidth{\columnwidth}
    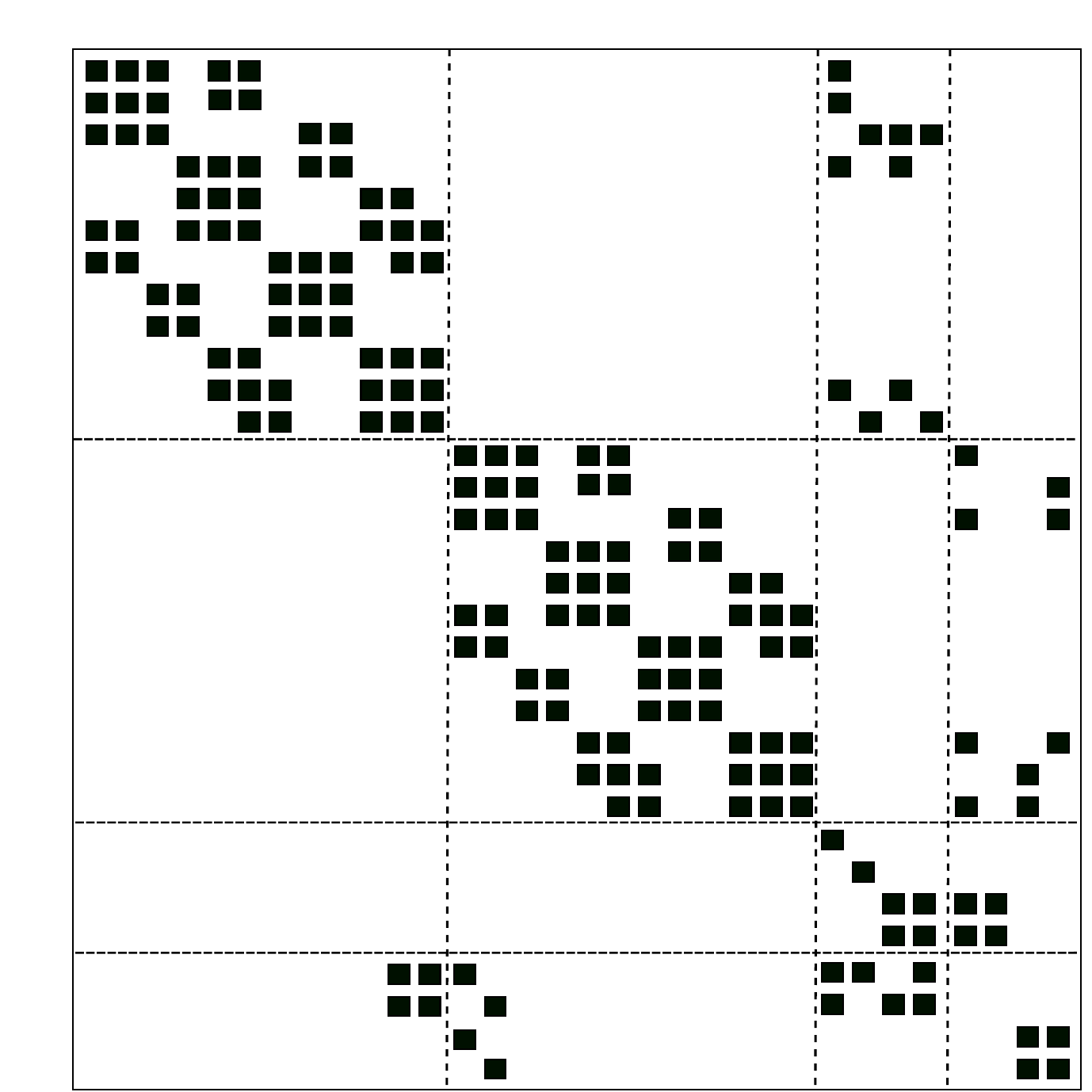

    \caption{%
    \addedaccepted{Schematic representation of the Jacobian matrix $J$ for the PDE
    $ \nabla^2 u(x,y) = S(x,y) $ and BC
    Eqs.~(\ref{eq_inner_bc_1}) and (\ref{eq_inner_bc_2}) when solved
    on the two subdomains of Fig.~\ref{fig_schur_grid}, with four by
    four points in each subdomain. Non-zero entries in $J$ are
    represented by small filled squares.
    The dashed lines delineate sub-matrices, analogous
    to the sub-matrices in Eq.~(\ref{eq_J_reordered}).
    Here $F_{11}$ and $F_{12}$ are empty since Eq.~(\ref{eq_inner_bc_2}) is
    imposed from the side of subdomains $\Omega_1$. On the other hand,
    $F_{21}$ and $F_{22}$ are non-empty as Eq.~(\ref{eq_inner_bc_1})
    is imposed from the side of subdomains $\Omega_2$.}
    }
    \label{fig_schematic_J}
\end{figure}

\addedaccepted{
As an example, suppose we wish to solve the elliptic equation
$ \nabla^2 u(x,y) = S(x,y) $ together with BC Eqs.~(\ref{eq_inner_bc_1}) 
and (\ref{eq_inner_bc_2}) on the 2-dimensional domain $\Omega$ that is divided
into two subdomains $\Omega_1$ and $\Omega_2$ as shown on the right of
Fig.~\ref{fig_schur_grid}.
Assume further that we cover each subdomains by a grid of four by four
points. Then, the subdomains share four points along the common interface
$\Gamma$ where the BCs are imposed. These BCs couple the equations on both
subdomains. The equations are then reordered following the
general structure given in Eq.~\ref{eq_J_reordered}.
After this reordering, the Jacobian coming from the equations has the
schematic form depicted in Fig.~\ref{fig_schematic_J}.
}
\subsection{Coordinate Patch}
\label{sec_grid}

Although equations in general relativity do not depend on
coordinate patches on the spacetime manifold, explicit coordinate patches
are needed when it comes to numerical-relativity calculations.
{\tt Elliptica} uses, but is not limited to, 
Cartesian and cubed spherical~\cite{Ronchi1996} coordinate systems.
In this section, we illustrate how the computational grid of a BHNS system
is covered.

The essence of the transformation between
Cartesian coordinates $x^i=(x,y,z)$
and cubed spherical coordinates $X^i= (X,Y,Z)$ is as 
follows~\cite{Tichy_2019}:
\begin{align}
        X &= \frac{x^I}{x^K},
        \nonumber \\
        Y &= \frac{x^J}{x^K},
        \nonumber \\
        Z &= \frac{x^K-r_{in}}{r_{out}-r_{in}},
        \label{eq_cubed_spherical_trans}
\end{align}
where, the indices $I, J, K \in \{1,2,3\}$ are all distinct, and
$X, Y \in[-1,1], Z\in[0,1]$. Here
\begin{align}
        r_{\text{in}} &= \frac{\sigma_{in}(X,Y)}{\sqrt{1+X^2+Y^2}},
        \nonumber \\
        r_{\text{out}} &= \frac{\sigma_{out}(X,Y)}{\sqrt{1+X^2+Y^2}},
        \label{eq_r_in-out}
\end{align}
in which, the inner boundary of the patch is determined by 
$\sigma_{\text{in}}(X,Y)$ and the outer boundary by 
$\sigma_{\text{out}}(X,Y)$.
Moreover, each $\sigma(X,Y)$ is related to Cartesian coordinates 
by the equation $\sigma(X,Y) = \sqrt{x^2+y^2+z^2}$.
For instance, if a perfect $2$-sphere is required as the boundary of a patch
then $\sigma(X,Y)= C$, similarly for a planar boundary
$\sigma(X,Y)= C\sqrt{1+X^2+Y^2}$, where $C$ is a constant.

To better capture the field falloff (expected to occur as powers of
$r^{-1}$ at large radii), we replace Z by,
\begin{align}
        \widetilde{Z} = \frac{\sigma_{\text{out}}}
                       {\sigma_{\text{out}}-\sigma_{\text{in}}}
        \left(1-\frac{\sigma_{\text{in}}}{r}\right),
\end{align}
in the outermost patches~\cite{Tichy_2019}.
Here $r=\sqrt{x^2+y^2+z^2}$ and still $\widetilde Z\in[0,1]$.
Lastly, in order to avoid coordinate singularities at the center of 
spheres, each spheroidal object contains a Cartesian coordinate patch 
around its center.

\begin{figure}[t]
	\includegraphics[width=1.0\linewidth]{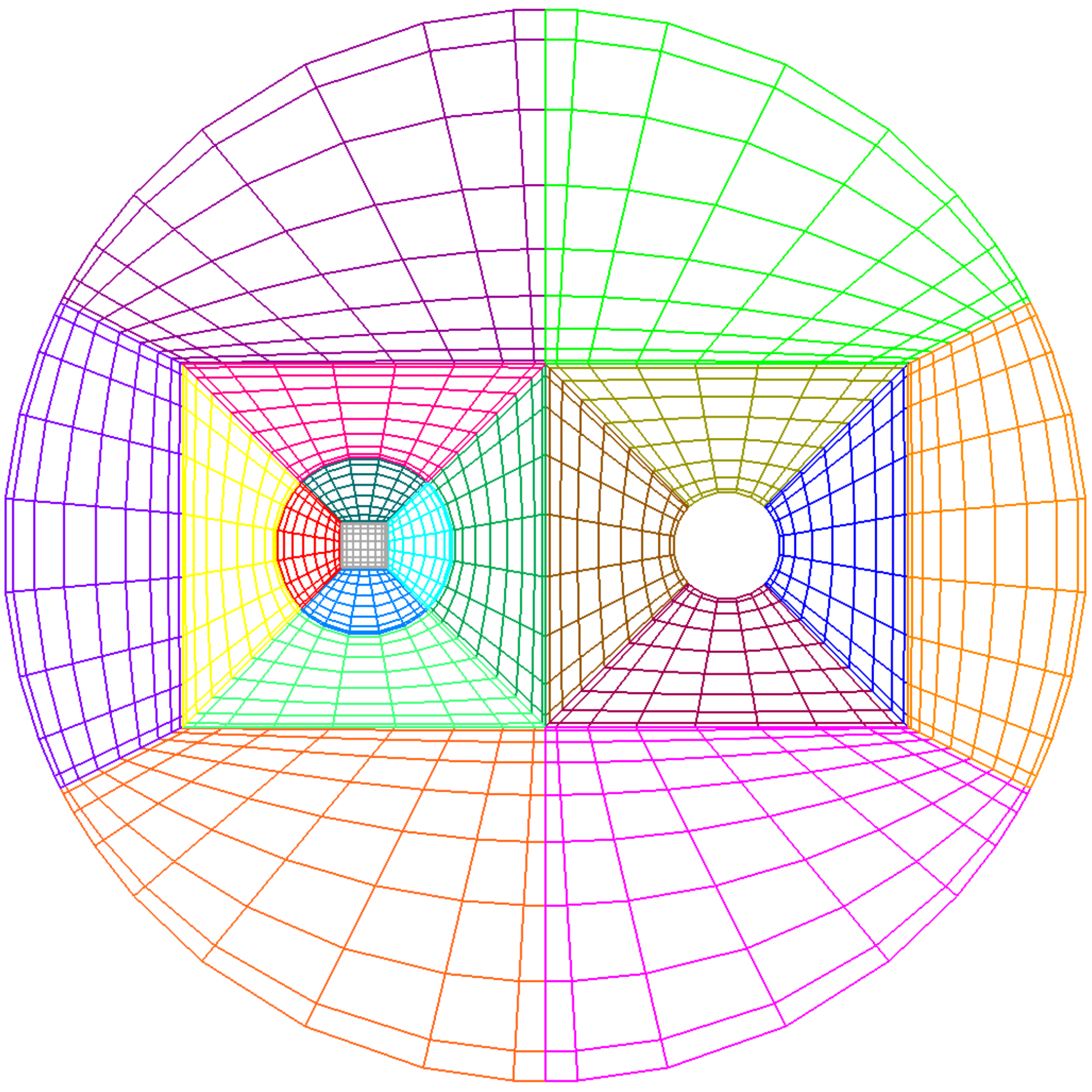}
	\caption{%
	An example of a computational grid
	that is using cubed spherical coordinates. Shown is the intersection
        of the domains with the $xy$-plane. The domains on the left
        cover a NS, the ones on the right cover the region around
        an excised BH.}
	\label{fig_full_cubed_sphere}
\end{figure}

Using these different maps, we can setup different patches 
with different surfaces for various needs, e.g. to cover the NS
or the space between the compact objects. 
These patches touch each other and never overlap.
They tile the entire computational domain, 
as depicted in Fig.~\ref{fig_full_cubed_sphere}.
We have also decided to increase the number of outermost 
patches in {\tt Elliptica} with respect to {\tt SGRID}~\cite{Tichy_2019}.
This leads to a more symmetric grid, higher angular resolution, and no need
for interpolation within interfaces
when setting up interface BCs, i.e., Eq.~(\ref{eq_inner_bc_1}) and 
Eq.~(\ref{eq_inner_bc_2}).

%
%
%┌───────────┐
%│ formalism │
%└───────────┘
\section{Formalism}
\label{sec_formalism}

In this section, we present the formalism applied to 
construct ID for quasi-equilibrium BHNSs. 
The ID of these systems are obtained
by solving Einstein's equations and Euler's equations.
In order to make these equations amenable to the numerical methods used
here, they are cast into elliptic
type PDEs. Specifically, we use the
extended conformal thin sandwich method 
(XCTS)~\cite{Pfeiffer_2003_2, York_1999}
for the Einstein's equations. For the fluid equations we use the
method described in~\cite{Tichy_2012}. Furthermore, 
there are two common approaches to treat BH singularities in
this context, the excision approach~\cite{Cook_2004} 
and the puncture approach~\cite{Brandt_1997}. 
{\tt Elliptica} uses the excision approach: the
BH singularity is excised from the computation domain and then
boundary conditions are imposed on the excised surface.
Below, we present the formulas and conventions used in {\tt Elliptica}.

Using the $3+1$ formalism (see e.g.~\cite{gourgoulhon_thebook}), 
we write the line element of a spacetime manifold as
\begin{equation}
        ds^2 = g_{\mu\nu}dx^\mu dx^\nu = 
        -\alpha^2 dt^2 + \gamma_{ij}(dx^i+\beta^i dt)(dx^j+\beta^j dt),
\end{equation}
where $g_{\mu\nu}$ is the pseudo-Riemannian metric of the 
spacetime manifold, $\alpha$ is the lapse function,
$\beta^i$ is the shift vector and 
$\gamma_{ij}$ is the induced metric of a spatial hypersurface $\Sigma_t$.
Using the normal vector $n^\mu$ orthogonal to $\Sigma_t$,
which can be interpreted as the $4$-velocity of an Eulerian observer, we have
$\gamma_{\mu\nu} = g_{\mu\nu} + n_\mu n_\nu$.
The extrinsic curvature on $\Sigma_t$ is defined by 
$
        K_{\mu\nu} = -\frac{1}{2} \pounds_{n} \gamma_{\mu\nu},
$
in which $\pounds_n$ is the Lie derivative along the normal vector.

The NS matter is described by a perfect fluid, therefore, 
the stress-energy tensor is
%┌──────────────────────┐
%│ stress energy tensor │
%└──────────────────────┘
\begin{align}
      T_{\mu\nu}= & 
           (\rho_0+\rho_0 \epsilon+P)u_\mu u_\nu + P g_{\mu\nu},
      \\
            = & \rho_0 h u_\mu u_\nu + P g_{\mu\nu},
\end{align}
where  $\rho_0$, $\epsilon$, $P$, $h$, and $u^\mu$ are, respectively,
the rest mass density, specific internal energy, pressure, 
specific enthalpy and the $4$-velocity of the fluid.

In order to employ the XCTS formalism, it is necessary to rescale 
the $3$-metric $\gamma_{ij}$ and decompose the extrinsic curvature 
$K^{ij}$ as follows:
\begin{align}
        \gamma_{ij} = & \psi^4 \bar \gamma_{ij},
        \\
        K^{ij} = & A^{ij}+\frac {1}{3} K \gamma^{ij}.
\end{align}
Here $\psi$ is the conformal factor, $\bar \gamma_{ij}$
the conformal 3-metric, $A^{ij}$ the traceless part of $K^{ij}$
and $K = \gamma_{ij}K^{ij}$.
Furthermore, it is convenient to project and then rescale the matter 
quantities as:
\begin{align}
        E= \, &n_\mu n_\nu T^{\mu\nu}  & = & \psi^{-6} \bar E,
        \\
        S= \, &\gamma^{ij}\gamma_{i\mu}\gamma_{j\nu}T^{\mu\nu} & =
           &\psi^{-6}\bar S,
        \\
        j^i = \, & -\gamma^{i}_{\mu} n_{\nu}T^{\mu\nu} &= &\psi^{-6}\bar j^i
        .
\end{align}
Moreover, in the XCTS formalism $A^{ij}$ is related to
the shift and the time derivative of conformal metric as follows:
\begin{align}
        A^{ij} &= \psi^{-10} \bar A^{ij},
        \\
        \bar A^{ij} &= \frac{1}{2 \bar \alpha} 
        \left((\bar L \beta)^{ij}-\bar \gamma^{ik}\bar 
	\gamma^{jl}\bar u_{kl}\right),
\end{align}
where
\begin{align}
        \bar u_{ij} &= \frac{\partial \bar \gamma_{ij}}{\partial t},
        \\
        (\bar L \beta)^{ij} &= \bar D^i \beta^j + \bar D^j \beta^i 
        - \frac{2}{3}\bar\gamma^{ij}\bar D_k \beta^k,
        \\
        \alpha &= \psi^6 \bar \alpha,
\end{align}
and $\bar D$ is the covariant derivative compatible 
with $\bar \gamma_{ij}$. 
Finally, the Einstein equations result in five coupled nonlinear
elliptic PDEs~\cite{York_1999, Pfeiffer_2003_2}:
%
%┌───────────┐
%│ equations │
%└───────────┘
\begin{align}
        \bar D^{2}\psi-\frac{1}{8}\psi \bar R-
        \frac {1}{12}\psi^{5}K^{2}+
        \frac{1}{8}\psi^{-7}\bar A_{ij}\bar A^{ij}
        \nonumber \\
        +2\pi \psi^{-1} \bar E &=& 0, \label{eq_psi}
        \\
        2\bar \alpha \left[\bar D_{j}(\frac {1}{2\bar \alpha}
  (\bar L \beta)^{ij})-\bar D_{j}(\frac {1}{2\bar \alpha }\bar u^{ij})
        -\frac {2}{3}\psi^{6}\bar D^{i}K\right]
        \nonumber \\
        -16\pi \bar \alpha \psi ^{4}\bar j^i &=&0,\label{eq_beta}
        \\
        \bar D^{2}(\bar \alpha \psi^{7})-
        (\bar \alpha \psi^{7}) \left[ \frac {1}{8}\bar R+\frac {5}{12}
        \psi^{4} K^{2}+\frac {7}{8}\psi^{-8}\bar A_{ij}\bar A^{ij}\right]
        \nonumber \\
        + \psi^{5}(\partial_{t} K-\beta^k \partial_{k} K)
        -2\pi \bar\alpha \psi^{5}(\bar E+2\bar S) &=&0,\label{eq_lapse}
\end{align}
where $\bar R$ is the Ricci scalar coming from $\bar \gamma_{ij}$.
These equation must be solved with appropriate boundary conditions 
at spatial infinity 
and on the surface of the excised region $H$
in order to have a unique physical solution. 
Before mentioning these boundary conditions, it is useful to decompose
$\beta^i$ such that in an inertial frame it has clearly identifiable
rotational and inspiral pieces at spatial infinity:
\begin{equation}
        \vec{\beta} =
        \vec{\beta}_0 +
        \vec{\Omega}_{\text{BHNS}}\times(\vec{r}-\vec {r}_{\text{CM}}) +
        \frac{v_r}{r_{\text{BHNS}}}(\vec{r}-\vec {r}_{\text{CM}}).
        \label{eq_shift_def}
\end{equation}
Here, $\vec{\Omega}_{\text{BHNS}}$ denotes the angular velocity 
of the BHNS system, 
$\vec {r}_{\text{CM}}$ is the center of mass
of the system, $v_r$ is the radial velocity,
$r_{\text{BHNS}}$ is the coordinate distance between the NS and BH's
centers, and $\vec{\beta}_0$ is determined by the elliptic
Eq.~(\ref{eq_beta}).
Finally, the following boundary conditions are imposed at spatial
infinity:
%┌─────┐
%│ B.C │
%└─────┘
\begin{equation}
  \label{eq_bc_inf}
  \lim_{r\to\infty}\psi = 1, \ \ \
  \lim_{r\to\infty}\beta^i_0 = 0, \ \ \
  \lim_{r\to\infty}\alpha\psi = 1.
\end{equation}
\addedaccepted{The boundary conditions in Eq.~(\ref{eq_bc_inf}) come from assuming
asymptotic flatness at spatial infinity, and furthermore going to
a frame that is rotating with $\vec{\Omega}_{\text{BHNS}}$ with respect
to the asymptotically flat inertial frame.
In order to have the excised surface $H$ corresponding to an
apparent horizon (zero expansion for outgoing null rays) in a state of
equilibrium, the following boundary conditions 
are imposed on}
$H$~\cite{Cook_2004}: 
\begin{align}
        \left\{
        \bar s^i\bar D_i \ln \psi+
        \frac{1}{4}\bar h^{ij}\bar D_i \bar s_j -
        \frac{K}{6}\psi^2+
        \frac{\psi^{-4}}{4}\bar A^{ij} \bar s_i \bar s_j
        \right\}_{H} &=&0,
        \label{eq_bc_psi_h}
        \\
        \left\{
        \beta^i - \alpha s^i-\epsilon_{ijk}
        \Omega^j_{\text{BH}}(x^k-\bar{x}^k_{\text{BH}})
        \right\}_{H}&=&0,
        \label{eq_bc_beta_h} 
        \\
        \left\{
        \bar s^i\bar D_i(\alpha \psi) 
        \right\}_{H} & = & 0,
        \label{eq_bc_alpha_h}
\end{align}
where, $s^i$ is the outward pointing unit normal on  $H$, 
$\bar s^i = \psi^2 s^i$, 
the induced metric on $H$ is $h_{ij}=\gamma_{ij}-s_i s_j$ and 
$\bar h^{ij} = \psi^4 h^{ij}$. 
Moreover, in Eq.~(\ref{eq_bc_beta_h}), 
$\epsilon_{ijk}$
is the totally anti-symmetric symbol and summation over repeated indices
is implied,
$\Omega^j_{\text{BH}}$ is a free vector to adjust the BH spin and
$\bar{x}^k_{\text{BH}}$ is the coordinate of the BH's center.

When an NS is present we also need matter equations. In order to solve the
fluid equations for an NS with an arbitrary spin, a purely spatial vector
spin (to encapsulate the rotational part of the fluid)
is introduced in~\cite{Tichy_2012} as follows:
\begin{equation}
        \label{eq_ns_spin_vector}
        w^i=\epsilon_{ijk}
        \Omega^j_{\text{NS}}(x^k-\bar{x}^k_{\text{NS}}),
\end{equation}
in which, $\bar{x}^k_{\text{NS}}$ denotes the coordinate of NS's center 
and $\Omega^j_{\text{NS}}$ is a vector related to the spinning motion of
the NS. Furthermore, assuming the BHNS is in 
a quasi-equilibrium state, we introduce 
an approximate helical time-like Killing vector 
$\xi^\mu$~\cite{Tichy_2012}. Consequently,
the fluid $4$-velocity can be projected along $\xi^{\mu}$ 
and a pure spatial vector $V^{\mu}$~\cite{Tichy_2012,Tichy_2016} as:
\begin{equation}
        u^{\mu} = 
        u^0(\xi^{\mu}+V^{\mu}) = 
        \frac{g^{\mu\nu}\frac{\partial \phi}{\partial x^{\nu}}+w^{\mu}}{h},
\end{equation}
where scalar $\phi$ encompasses the irrotational part of the fluid.
Using the approximations described in~\cite{Tichy_2012,Tichy_2016,Tichy_2019} the
Euler equation becomes an elliptic PDE for $\phi$
%
%┌──────────────┐
%│ phi equation │
%└──────────────┘
\begin{align}
	& \frac{c\left(\rho_0\right)\alpha}{h} \psi^{-4} \bar\gamma^{ij} 
	\partial_i \partial_j \phi
	 - \frac{\rho_0\alpha}{h} \psi^{-4} \bar\gamma^{ij}  
	\bar \Gamma^k_{ij}\partial_k \phi
	\nonumber \\
	&
	+ 2\frac{\rho_0\alpha}{h} \psi^{-5} 
	\bar \gamma^{ij}(\partial_i\psi) (\partial_j\phi)
	+ \left(D_i \frac{\rho_0 \alpha}{h}\right) \left(D^i \phi\right)
	\nonumber \\
	& D_i \left[ \frac{\rho_0 \alpha}{h}  w^i
        -\rho_0 \alpha u^0 (\beta^i + \xi^i) \right] = 0.
	\label{eq_phi}
\end{align}
Here, following~\cite{Tichy_2019}
\begin{equation}
c(\rho_0) = \rho_0 + \epsilon \rho_{0c}
\left(\frac{\rho_{0c}-\rho_0}{\rho_{0c}}\right)^4 ,
\end{equation}
$\rho_{0c}$ is rest mass density at the NS's center,
and $\epsilon$ is a small number (generally $0.1$).
\addedaccepted{On the NS's surface we have $\rho_0 = 0$, hence, the following
boundary condition is imposed on the surface~\footnote{In practice we
use normal vector on the NS's surface instead of $D_i\rho_0$}}
%┌─────────┐
%│ phi B.C │
%└─────────┘
\begin{equation}
       D^{i}\phi D_i\rho_0+w^{i} D_{i} \rho_0-
       h u^0(\beta^i+\xi^i) D_i\rho_0 = 0.
\end{equation}

One more equation is needed to close the system of Einstein-Euler
equations. This equation is the equation of state (EoS) for NS's matter. 
In this work, we use a piecewise polytropic EoS.
The pieces valid between the densities $\rho_1,\rho_2,\hdots,\rho_n$
~\cite{Read_2009} are written as
\begin{align}
        P=
        \begin{cases}
                K_0 \rho_0^{\Gamma_0} \;& \rho_0 \leq \rho_1
                \\
                K_1 \rho_0^{\Gamma_1} \;& \rho_1 \leq \rho_0 \leq \rho_2
                \\
                \vdots
                \\
  K_{n-1} \rho_0^{\Gamma_{n-1}} \;& \rho_{n-1} \leq \rho_0 \leq \rho_{n},
        \end{cases}
\end{align}
in which, $\Gamma_i$ denotes the polytropic exponent and  
$K_i$ the polytropic constant. Moreover, one can write the 
rest mass density $\rho_0$, pressure $P$ and  specific internal energy  
$\epsilon$ in terms of specific enthalpy  $h$:
 \begin{align}
         \rho_0(h) &= K_i^{-n_i}\left(\frac{h-1-a_i}{n_i+1}\right)^{n_i},
        \nonumber\\
         P(h) &= K_i^{-n_i}\left(\frac{h-1-a_i}{n_i+1}\right)^{n_i+1},
        \nonumber\\
         \epsilon(h) &= \frac{a_i+n_i(h-1)}{n_i+1},
 \end{align}
where $n_i=\frac{1}{\Gamma_i-1}$ is the polytropic index and $a_i$'s are
constants which ensure the continuity of EoS:
\begin{align}
        a_0 &= 0, 
        \nonumber \\
        a_i &=
        a_{i-1}+
        \frac{K_{i-1}}{\Gamma_{i-1}-1}\rho_{i}^{\Gamma_{i-1}-1}
        -\frac{K_i}{\Gamma_i-1}\rho_i^{\Gamma_i-1}.
\end{align}
Lastly, the specific enthalpy $h$ is determined by an algebraic equation
in terms of metric variables and fluid velocities
~\cite{Tichy_2012}:
%
%┌──────────┐
%│ enthalpy │
%└──────────┘
\begin{align}
        h &=& \sqrt{L^2 - (D_i \phi + w_i)(D^i \phi + w^i)},
        \nonumber \\
        L^2 &=& \frac{b + \sqrt{b^2 - 4\alpha^4 [(D_i \phi + w_i) w^i]^2}}
              {2\alpha^2},
        \nonumber \\
        b &=& [ (\xi^i+\beta^i)D_i \phi - C]^2 + 
          2\alpha^2 (D_i \phi + w_i) w^i ,
        \label{eq_enthalpy}  
\end{align}
in which $C$ denotes a constant of integration that determines the 
baryonic mass of the NS. We observe that 
the specific enthalpy connects metric variables and matter variables,
in the other words, macro-physics and micro-physics.
\addedaccepted{In the next section, we explain how the solution of these 
elliptic equations and free parameters (such as the $C$ above) are found
in order to construct proper ID.}

%
%┌──────────────────┐
%│ numerical method │
%└──────────────────┘
\section{Numerical Method}
\label{sec_numerical_method}
In this section, we demonstrate the main iteration algorithm 
for the construction of ID for a BHNS system. 
Input parameters for {\tt Elliptica}
are the baryonic mass $M_{\text{B}}$, the EoS, and the angular velocity 
$\Omega^j_{\text{NS}}$ in Eq.~(\ref{eq_ns_spin_vector}) 
for the spin vector of the NS. Moreover, we can specify
the irreducible mass $M_{\text{irr}}$ and dimensionless spin $\chi^j$
of the BH and also the coordinate distance between the centers of the
NS and BH, as well as the orbital angular velocity and the radial velocity
of the system. For the free data, we currently use:
%┌───────────┐
%│ free data │
%└───────────┘
\begin{align}
        \bar \gamma_{ij} &= \delta_{ij},
        \label{eq_free_data_gamma}
        \\
        K & = 0,
        \label{eq_free_data_K}
        \\
        \bar u_{ij} & = 0.
\end{align}
To account for quasi-equilibrium we also set $\partial_t K = 0$.
Since our numerical method is iterative, we need an initial guess
for the fields $\psi$, $\beta^i$, $\alpha$, $\phi$, and $h$
on the computational grid. We use a superposition of well known 
analytic solutions for single objects for this guess. 
For the star we use a Tolman-Oppenheimer-Volkoff (TOV) solution and
for the BH a Schwarzschild solution in isotropic coordinates.
The initial value of the enthalpy $h$ then is computed from the
TOV star and the irrotational velocity potential is set to
$\phi = -\Omega_{\text {BHNS}}(y_{\text {NS}}-y_{\text{CM}})x$, 
where $y_{\text {NS}}$ is the $y$-coordinate of NS's center 
and $y_{\text{CM}}$ is the $y$-coordinate of the system's center of mass
(note the objects are centered on the $y$-axis). 

Having determined the initial guess, 
we can now use the Newton-Raphson algorithm~(\ref{algo_newton})
to solve the pertinent coupled elliptic 
PDEs~(\ref{eq_psi}), (\ref{eq_beta}), (\ref{eq_lapse}) and (\ref{eq_phi}). 
However, 
\addedaccepted{due to the presence of the matter~(NS) and the use of the XCTS formalism,}
several obstacles need to be overcome before we can find a solution with the correct properties%
\addedaccepted{, e.g., see ~\cite{Tacik2016}.}
First, the surface of the NS star
is not known in advance and is changing at each iteration. 
As a result, one should find this surface and then adjust 
the coordinate patches such that the surface is a patch boundary. 
Otherwise spectral convergence cannot be achieved since the
matter fields are not smooth across the star surface.
Second, the NS's center and mass start drifting from the desired values which
usually causes instabilities.
Third, BH's mass and spin deviate from the target values which
results in a solution with the wrong physical properties.
Finally, the ADM momentum of the system grows which gives rise to 
instabilities in our iterative procedure, and will also impart a kick
on the system's center of mass visible during subsequent
evolution. Such a drift can also cause undesirable
coordinate effects when extrapolating gravitational waves 
to infinity. 

Therefore, it is crucial to monitor, adjust and control various parameters
and values at each step of the Newton-Raphson algorithm~(\ref{algo_newton})
in order to construct ID with the correct properties.

\subsection{Diagnostics}
\label{sec_diagnostics}
As mentioned in the previous subsection, 
while we are solving the coupled elliptic 
Eqs.~(\ref{eq_psi}), (\ref{eq_beta}), (\ref{eq_lapse}) and (\ref{eq_phi}),
there is no guarantee that the mass and spin of BH or the mass of NS 
reach the target values specified in the parameter file. Moreover, the
ADM momentum of the system generally does not vanish.
In this section, we show the formulas used to calculate spins, masses, 
ADM momenta and angular momenta for a BHNS system.
We also explain how we attain the desired values for these quantities.

Starting with the NS, the baryonic mass density $4$-current
is defined~\cite{gourgoulhon_thebook}:
\begin{equation}
       J^{\mu}_B = \rho_0 u^{\mu},
\end{equation}
thus, the baryonic mass density as measured by an Eulerian observer is
$-J^{\mu}_B n_{\mu}$ and the baryonic mass is:
%┌─────────────┐
%│ Baryon mass │
%└─────────────┘
\begin{equation}
        M_B = \int_{\text{NS}} -J^{\mu}_B n_{\mu} dV.
        \label{eq_baryonic_mass}
\end{equation}
In {\tt Elliptica} we write Eq.~(\ref{eq_baryonic_mass}) 
in terms of the $3+1$ decomposition formalism~\cite{gourgoulhon_thebook} 
to compute the baryonic mass:
\begin{equation}
        M_B = \int_{\text{NS}} \rho_0 \alpha 
              \psi^6 \sqrt{\bar{\gamma}} d^3x,
        \label{eq_baryonic_mass_simplified}
\end{equation}
where $\bar{\gamma}$ is the determinant of $\bar{\gamma}_{ij}$ and the
integration is taken over the volume of the NS.
We note that $\rho_0 = \rho_0(h)$ and as shown in Eq.~(\ref{eq_enthalpy}) 
the enthalpy depends on a constant $C$, i.e., $h = h(C)$, which implies
$\rho_0 = \rho_0(C)$. So by adjusting $C$ we can keep the baryonic mass 
constant at each step of the iteration.
%
%┌───────────┐
%│ BH masses │
%└───────────┘

For the BH, we calculate two masses, the irreducible mass $M_{\text{irr}}$
and the Christodoulou mass $M_{\text{Chr}}$ 
which are defined as follows respectively:
\begin{align}
        M_{\text{irr}} &= \sqrt{
        \frac{\oint_{\text{AH}} dA}{16\pi}},
        \label{eq_irr_mass}
        \\
        M_{\text{Chr}} & = \sqrt{
                M_{\text{irr}}^2 +
        \frac{S^2_{\text{BH}}}{4 M_{\text{irr}}^2}},
        \label{eq_chris_mass}
\end{align}
in which $dA$ is the proper surface element of the apparent horizon 
defined in {\tt Elliptica} as
\begin{equation}
       dA = \sqrt{\gamma_{ij}
       \frac{\partial x^i}{\partial y^a}
       \frac{\partial x^j}{\partial y^b}}
       dy^a dy^b,
\end{equation}
where $y^a$ are coordinates on the apparent horizon and
$S_\text{BH}$ is the spin of the BH which will be defined shortly. 
%%%%%
The apparent horizon is a perfect $2$-sphere in the coordinates used,
owing to the boundary conditions on the BH, i.e.,
Eqs.~(\ref{eq_bc_psi_h}, \ref{eq_bc_beta_h}, and \ref{eq_bc_alpha_h}). 
Thus, by adjusting the radius of the apparent horizon, 
we can drive the irreducible mass to the value prescribed 
in the parameter file.
%
%┌────────────────┐
%│ BH and NS Spin │
%└────────────────┘

For the BH's spin, we use the flat space coordinate rotational
Killing vector following~\cite{Campanelli_2007} on the apparent horizon:
\begin{align}
        \vec{\phi}_x & = -(z-z_c)\vec{\partial}_y+(y-y_c)\vec{\partial}_z,
        \nonumber
        \\
        \vec{\phi}_y & = +(z-z_c)\vec{\partial}_x+(x-x_c)\vec{\partial}_z,
        \nonumber
        \\
        \vec{\phi}_z & = -(y-y_c)\vec{\partial}_x+(x-x_c)\vec{\partial}_y,
\end{align}
in which $(x_c,y_c,z_c)$ is the coordinate center of the BH
and $(\vec{\partial}_i)_{i \in \{1,2,3\}}$ are the basis vectors associated
with the coordinates used.
The following integral over the apparent horizon (AH) yields
the spin of the BH:
\begin{equation}
        S_i = 
        \frac{1}{8\pi}\oint_{\text{AH}} (\vec{\phi}_i)^j s^k K_{jk}dA .
        \label{eq_evaluate_bh_spin}
\end{equation}
The dimensionless spin is defined by
\begin{equation}
        \chi_i := \frac{S_i}{M^2_{\text{Chr}}}.
\end{equation}
To adjust the value of the spin to the target value we note 
that $K^{ij} = K^{ij}(\vec{\beta})$ on the apparent horizon and  
$\beta^i = \beta^i(\vec{\Omega}_{\text{BH}})$ by Eq.~(\ref{eq_bc_beta_h});
thus by adjusting $\Omega^i_{\text{BH}}$ the spin is controlled.

For the NS spin, we have two options. 
First we can use the method described in~\cite{Tichy_2019},
\begin{equation}
        S_i = J_i-\epsilon_{ijk} R^j_c P_k,
	\label{eq_JRP}
\end{equation}
where $J_i$, $R^i_c$, and $P_i$ are, respectively, 
the angular momentum, center and momentum of the NS 
defined in~\cite{Tichy_2019}. 
The second option is to evaluate Eq.~(\ref{eq_evaluate_bh_spin}) 
on the NS's surface.
Note that from $\Omega^j_{\text{NS}}$ in Eq.~(\ref{eq_ns_spin_vector})
we cannot directly infer the NS spin, but the values of
$\Omega^j_{\text{NS}}$ for various spins can be
found in~\cite{Tacik2016, Tichy_2019, Papenfort:2021hod}.

%┌────────────────────────────────┐
%│ ADM linear and angular momenta │
%└────────────────────────────────┘
%
Since the chosen free data $\bar{\gamma}_{ij}$ in 
Eq.~(\ref{eq_free_data_gamma}) satisfies the
quasi-isotropic gauge condition~\cite{gourgoulhon_thebook} and since
$K$ in Eq.~(\ref{eq_free_data_gamma}) meets the
asymptotic maximal gauge condition~\cite{gourgoulhon_thebook}, 
the ADM linear momenta and angular momenta of the system can be 
defined~\cite{gourgoulhon_thebook} as follows:
\begin{align}
        P^{\infty}_i  & = \frac{1}{8\pi} \lim_{S_t \to \infty} 
        \oint_{S_t} (K_{jk}-K \gamma_{jk})(\vec{\partial}_i)^j s^k dA,
        \label{eq_linear_momentum}
        \\
        J^{\infty}_i  & = \frac{1}{8\pi} \lim_{S_t \to \infty} 
        \oint_{S_t} (K_{jk}-K\gamma_{jk})(\vec{\phi}_i)^j s^k dA,
        \label{eq_angular_momentum}
\end{align}
and 
\begin{align}
        \vec{\phi}_x & = 
         -(z-z_{\text{CM}})\vec{\partial}_y+
         (y-y_{\text{CM}})\vec{\partial}_z,
        \nonumber
        \\
        \vec{\phi}_y & = 
        +(z-z_{\text{CM}})\vec{\partial}_x+
        (x-x_{\text{CM}})\vec{\partial}_z,
        \nonumber
        \\
        \vec{\phi}_z & = 
        -(y-y_{\text{CM}})\vec{\partial}_x+
        (x-x_{\text{CM}})\vec{\partial}_y.
\end{align}
We note that $K_{ij}$ is a function of $\vec{\beta}$, while 
$\vec{\beta}$ itself is a function of $\vec{r}_{\text{CM}}$ 
through Eq.~(\ref{eq_shift_def}). Therefore, by adjusting the
freely specifiable parameter $\vec{r}_{\text{CM}}$, $P^{\infty}_i$ 
can be driven to zero during the solve.

Lastly, to calculate the total ADM mass of the system, we use the following%
~\cite{gourgoulhon_thebook}:
\begin{align}
\label{eq_adm_mass}
         &M_{ \text{ADM}} = M_{\text{H}} +
	\nonumber
        \\
	&\int_{\Sigma_t} \left[ \psi^5 E 
        + \frac{1}{16\pi} \left( \bar A_{ij} \bar A^{ij} \Psi^{-7}
        - {\bar R} \psi - \frac{2}{3} K^2 \psi^5 \right) \right]
        \sqrt{\bar \gamma} d^3 x.
\end{align}
Here, $\Sigma_t$ is the spatial hypersurface where the ID 
are constructed and $\bar \gamma$ is the determinant of $\bar \gamma_{ij}$.
$M_{\text{H}}$ is defined as
\begin{align}
	M_{\text{H}} := 
	- \frac{1}{2\pi} \oint_{\text{H}}
        {\bar s}^i \bar D_i \psi \, \sqrt{\bar h}\, d^2 y ,
\end{align}
where $\bar h := \text{det}(\bar h_{ij})$ and $\bar h_{ij}$ is 
the induced metric by $\bar \gamma_{ij}$ on the excised surface H.

In summary, we use the free parameters of the system, for instance,
$C$ and $\vec{r}_{\text{CM}}$ among others,
to obtain the requested physical properties of the system. 
In the next subsection, we show how to adjust these parameters (in a slow and
smooth way) to reach a stable solution.

\subsection{Iteration Algorithm}
\label{sec_iteration_algorithm}

Having set the initial fields and free data, the next step is to refine
the answers. Loosely speaking, the overall procedure is to start 
at low resolution and to keep solving the PDEs and 
adjusting the parameters until the error is below a desired tolerance; 
then, increase the resolution as often as needed and solve
again (still adjusting the parameters).
The detailed explanation of this iteration scheme to find a stable,
unique and physical solution of Einstein-Euler equations is as follows:
%
%┌─────────────────────┐
%│ iteration algorithm │
%└─────────────────────┘

  1. Solve each elliptic 
  Eqs.~(\ref{eq_psi}), (\ref{eq_beta}), (\ref{eq_lapse}) and (\ref{eq_phi})
  one after another using the Newton-Raphson method. 
  For each equation, only one step is taken in the method
  \addedaccepted{\footnote{Please note that we do not solve all coupled equations 
  at once, i.e., each elliptic equation is solved separately one after another.
  In other words, while solving each equation for the field, the other
  coupled fields are treated like fixed source terms.
  However, since we use an iterative procedure we eventually find the
  solution to the complete system of coupled equations.}}
  
  2. Update the field values 
  $\Psi = \{\phi,\psi,\alpha\psi,B^i\}$ using the
  relaxation scheme 
  $\Psi = \lambda \Psi_{\text{new}} + (1-\lambda) \Psi_{\text{old}}$,
  where $\Psi_{\text{new}}$ is the solution just found by the 
  Newton-Raphson method and $\Psi_{\text{old}}$ is its previous value
  (usually $\lambda = 0.2$%
  \addedaccepted{~\footnote{To determine the value of the relaxation parameter
  $\lambda$ here, we have conducted numerical experiments, with the goal
  of finding a $\lambda$ such that our iteration algorithm converges
  quickly enough. Below we use this kind of relaxation also when we update
  or adjust other quantities, albeit with different values.}}).

  3. Adjust $\Omega^i_{\text{BH}}$ to reach the target value 
  $\chi^i_*$ for the dimensionless BH's spin using
\begin{equation} 
  \Omega^i_{\text{BH,new}} = 
  \Omega^i_{\text{BH,old}}+\lambda\Delta \chi^i \Omega_{\text{BHNS}},
\end{equation}
  where $\Delta \chi^i = \chi^i_* - \chi^i$ and $\chi^i$ is the current
  value of dimensionless spin, and $\lambda$ is usually set to $0.3$.

  4. Adjust the excision radius of the BH to reach the target
  value for the irreducible mass. 
  The new radius is
\begin{equation}
  r_{\text{new}} = r_{\text{old}}(1+\lambda \frac{\Delta M}{M_*}).
\end{equation}
  Here $\Delta M = M_* - M$, $M_*$ is the target value, $M$
  is the current value of the irreducible mass of the BH, and
  $\lambda$ is generally set to $0.3$.

  5. Find the constant $C$ in Eq.~(\ref{eq_baryonic_mass_simplified})
  to achieve the prescribed value for baryonic mass of the NS.

  6. Adjust $\vec{r}_{\text{CM}}$ to drive the linear ADM momentum 
  of the BHNS system to zero. The linear momentum in the $z$-direction 
  is very small 
  ($\frac{|P^{\infty}_z|}{M_{\text{ADM}}} < 10^{-9}$)
  therefore we only need to adjust 
  $x_{\text{CM}}$ and $y_{\text{CM}}$ as follows:
  \begin{align}
          x_{\text{CM,new}} & = 
          x_{\text{CM,old}} + \lambda
          \frac{P^{\infty}_y}{\Omega_{\text{BHNS}} M_{\text{ADM}}},
          \nonumber
          \\
           y_{\text{CM,new}} & = 
          y_{\text{CM,old}} - \lambda
          \frac{P^{\infty}_x}{\Omega_{\text{BHNS}} M_{\text{ADM}}},
  \end{align}
  in which, $M_{\text{ADM}}$ is the total ADM mass of the system;
  $\lambda$ generally is set to $0.2$.

  7. Update the enthalpy \addedaccepted{in the patches that cover the NS interior} 
  using Eq.~(\ref{eq_enthalpy}) with the relaxation method
  $h = \lambda h_{\text{new}}+(1-\lambda)h_{\text{old}}$; here,
  $\lambda$ is usually set to $0.1$.

  8. If we want to also determine $\Omega_{\text{BHNS}}$ (and not just
  use a given value for it), we use the force balance 
  Eq.~(\ref{eq_force_balance}).
  Specifically, we find $\Omega_{\text{BHNS}}$
  such that the following holds at the NS's center~\cite{Tichy_2016}:
  \begin{align}
          \partial_i &\ln[\alpha^2 - 
          (\beta^i+\xi^i+\frac{w^i}{hu^0})
          (\beta_i+\xi_i+\frac{w_i}{h u^0})] = 
          -2\partial_i \ln \Gamma,
          \nonumber
          \\
          \Gamma & = 
          \frac{\alpha u^0 [1-(\beta^i+\xi^i+\frac{w^i}{hu^0})
            \frac{D_i \phi}{\alpha^2 h u^0}-
            \frac{w_i w^i}{(\alpha^2 h u^0)^2}]}
            {\sqrt{1-(\beta^i+\xi^i+\frac{w^i}{hu^0})(\beta_i+\xi_i
                  +\frac{w_i}{h u^0})
                  \frac{1}{\alpha^2}}} .
          \label{eq_force_balance}
  \end{align}
  Here $\partial_i = \frac{\partial}{\partial x^i}$ and 
  $\Gamma$ is kept fixed. We use this along the line connecting
  the centers of the two objects.

  9. Extrapolate matter fields $\phi$, $w^i$, and $h$ outside 
  the NS. \addedaccepted{This serves two purposes. First, a smoothly extrapolated
  $h$ helps with step $11$, where we use a root finder to update the star
  surface location.
  Second, to interpolate the matter fields to the new grid (of step $12$)
  we need the values of $\phi$ and $w^i$, even outside the star,
  if the NS surface expands in step $11$.}
  To extrapolate $w^i$ outside, we apply Eq.~(\ref{eq_ns_spin_vector}).
  For the fields $\phi$ and $h$, 
  at each collocation point on the NS's surface
  with coordinate radius $r_0$, we extrapolate them using:
  \begin{equation}
	\label{eq_ns_extrap}
         f(r) = \left(a+\frac{b}{r}\right)
                \exp\left(-c_0\frac{r}{r_0}\right),
  \end{equation}
  where the coefficients $a$ and $b$ are found by demanding 
  $C^1$ continuity across the surface, 
  $c_0$ is generally set to $0.01$, and $r$ is the coordinate distance
  from the star's center.%
  \addedaccepted{ Eq.~\ref{eq_ns_extrap} damps the matter fields exponentially
  so that the NS's surface cannot expand too much in one iteration.
  We have found that this particular, continuous
  extrapolation function works best within the iterative solve.}

  10. To avoid the drifting of the NS's center located at $\vec r_0$ we
  shift the NS matter (given by $h$) to keep the star's center
  fixed. Using a Taylor expansion (for small shifts) we find that $h$
  needs to be updated as follows:
  \begin{equation}
          h_{\text{new}}(\vec{r}) = 
            h_{\text{old}}(\vec{r}) - 
            (\vec r - \vec r_0)\cdot\vec{\nabla} h_{\text{old}}(\vec{r}).
  \end{equation}
  
  11. Find the new surface of the NS. Using a root finder we find $r$
  such that 
  \begin{equation}
          h(r)- 1 = 0,
  \end{equation}
  where $r$ is the distance from the NS's center.
  This yields the new location of the NS's surface in
  spherical coordinates $(r,\theta,\phi)$
  \addedaccepted{and thereby determines the pertinent $\sigma(X,Y)$ introduced 
   in Sec.~\ref{sec_grid}.}

  12. If the resolution is changed or the surface of NS or BH has 
  been changed, create a new grid and set the fields values using 
  a spectral interpolation from the previous grid. 
  At this step a new grid with the new surface fitting patches for
  the NS and BH surface are created and the values of the fields are 
  interpolated from the previous grid.

  13. Exit, if all the iterations at all requested resolutions have reached
  our error criterion, otherwise proceed to step $1$.
  The criterion we use to exit is when 
  Hamiltonian and momenta constraints, 
  Eqs.~(\ref{eq_ham_constraint}) and (\ref{eq_mom_constraint}),
  are no longer decreasing (because they have reached the truncation error
  for the resolution).

To give some context, 
the number of outermost iterations for this BHNS system
is generally about $250$ at the lowest resolution 
(generally $12$ points at each direction in each patch). It decreases when
increasing the resolution and ends up at about $50$ at the highest
resolution (generally $20$ points at each direction per patch).
The maximum resolution itself is determined by the
maximum constraint violations we are willing to tolerate.
\addedaccepted{Using shared-memory multiprocessing,}
the ID computation for this configuration generally takes 
$\sim104$ hours \addedaccepted{of actual wall clock time}
on a single Intel Xeon node with $20$ cores of FAU's KOKO cluster.
A summary of this iterative scheme is shown in 
algorithm~(\ref{algo_main_iteration}).

\begin{figure}[t]
\begin{algorithm}[H]
\caption{main iteration scheme to find the ID for a BHNS system}
\label{algo_main_iteration}
\begin{algorithmic}[1]
\For{ each resolution }        
\While{ constraints have not plateaued yet}
  \State Solve \addedaccepted{ the elliptic}
Eqs.~(\ref{eq_psi}), (\ref{eq_beta}), (\ref{eq_lapse}), and (\ref{eq_phi})
\addedaccepted{one after another};
  \State Update the fields being solved for using the relaxation scheme
         $\Psi = \lambda \cdot \Psi_{\text{new}} + (1-\lambda) \cdot \Psi_{\text{old}}$;
   \label{step_relax}
  \State Adjust the BH parameter $\Omega^i_{\text{BH}}$ 
  to satisfy the target dimensionless spin;
  \State Adjust the BH radius to satisfy the target irreducible mass;
  \State Find Euler's equation constant $C$ to fix the NS 
  baryonic mass;
  \State Adjust the system's center of mass $\vec{r}_{\text{CM}}$
  to drive the ADM linear momenta to zero;
  \State Update the enthalpy using Eq.~(\ref{eq_enthalpy}) with 
	the relaxation method 
	$h = \lambda h_{\text{new}}+(1-\lambda)h_{\text{old}}$.
  \State If desired, use the force balance Eq.~(\ref{eq_force_balance})
  to adjust $\Omega_{\text{BHNS}}$.
  \State Extrapolate matter fields $\phi$, $w^i$ and $h$ to the outside 
  of the NS. 
  \State Shift the matter to keep the NS's center fixed.
  \State Find the new location of NS surface;
  \State If needed, create a new grid and interpolate the fields values 
  from the previous grid.
\EndWhile
\EndFor
\end{algorithmic}
\end{algorithm}
\end{figure}

%┌─────────┐
%│ results │
%└─────────┘
\section{Results}
\label{sec_results}
\addedaccepted{In this section, we present several validations of our implementation
and provide first proof-of-principle dynamical simulations.
In sec \ref{sec_spectral_convergence_test} we test the 
spectral convergence of the code for a non-spinning BHNS system
(named {\tt SXS1}) and for a system (named {\tt S1S2}),
where both BH and NS have high spins in arbitrary directions.
In sec \ref{sec_quasi_eq} we construct 
different sets of ID to confirm the validity of the data when
comparing to analytical approximations. We further push
the NS spin close to breakup value.
Finally in sec \ref{sec_evolution_tests} we evolve the ID
(summarized in Tab.~\ref{tab_runs}) using the
{\tt BAM} code~\cite{Brugmann2007,Thierfelder_2011,Dietrich_2015,%
Bernuzzi_2016,Chaurasia2021},
to check that the binaries behave as expected. 
In all configurations, the EoS is polytropic with $K=92.12$ and $\Gamma=2$
unless otherwise is mentioned.
}

\begin{table}[!h]
\centering
\caption{\addedaccepted{The ID that are evolved via the BAM code in this work.
We have listed the irreducible mass $M_{\text{irr}}$, baryonic mass
$M_{\text{B}}$,
the mass ratio of the binary $q$, the initial coordinate distance $s$,
the total dimensionless spin of each object $\chi_{\text{BH}}$,
$\chi_{\text{NS}}$, and the EoS. 
Here K92 denotes a polytropic EoS with $K=92.12$ and $\Gamma=2$.
The {\tt S1S2} ID 
has $\vec \chi_{BH} = (-0.46,-0.46,-0.46)$ and 
$\vec \chi_{NS} = (0,0.32,0.32)$.} }
\label{tab_runs}
\renewcommand{\arraystretch}{1.5}
\begin{tabular}{*{8}{>{\centering\arraybackslash}p{.11\linewidth}}}
\hline \hline 
Name       & $M_{\text{irr}}$ & $M_{\text{B}}$ & q  & s    & $\chi_{BH}$ & $\chi_{NS}$ & EoS \\ \hline 
{\tt SXS1} &   $8.4$            &  $1.4$    & $6.5$ & $82$ &  $0.0$      & $0.0$       & $K92$  \\        
{\tt S1S2} &   $5.2$            &  $1.4$    & $4$   & $56$ &  $0.8$      & $0.45$      & $K92$   \\ \hline \hline 
\end{tabular}
\end{table} 

\subsection{Spectral Convergence Test}
\label{sec_spectral_convergence_test}
\begin{figure}[t]
    \includegraphics[width=1.0\linewidth]{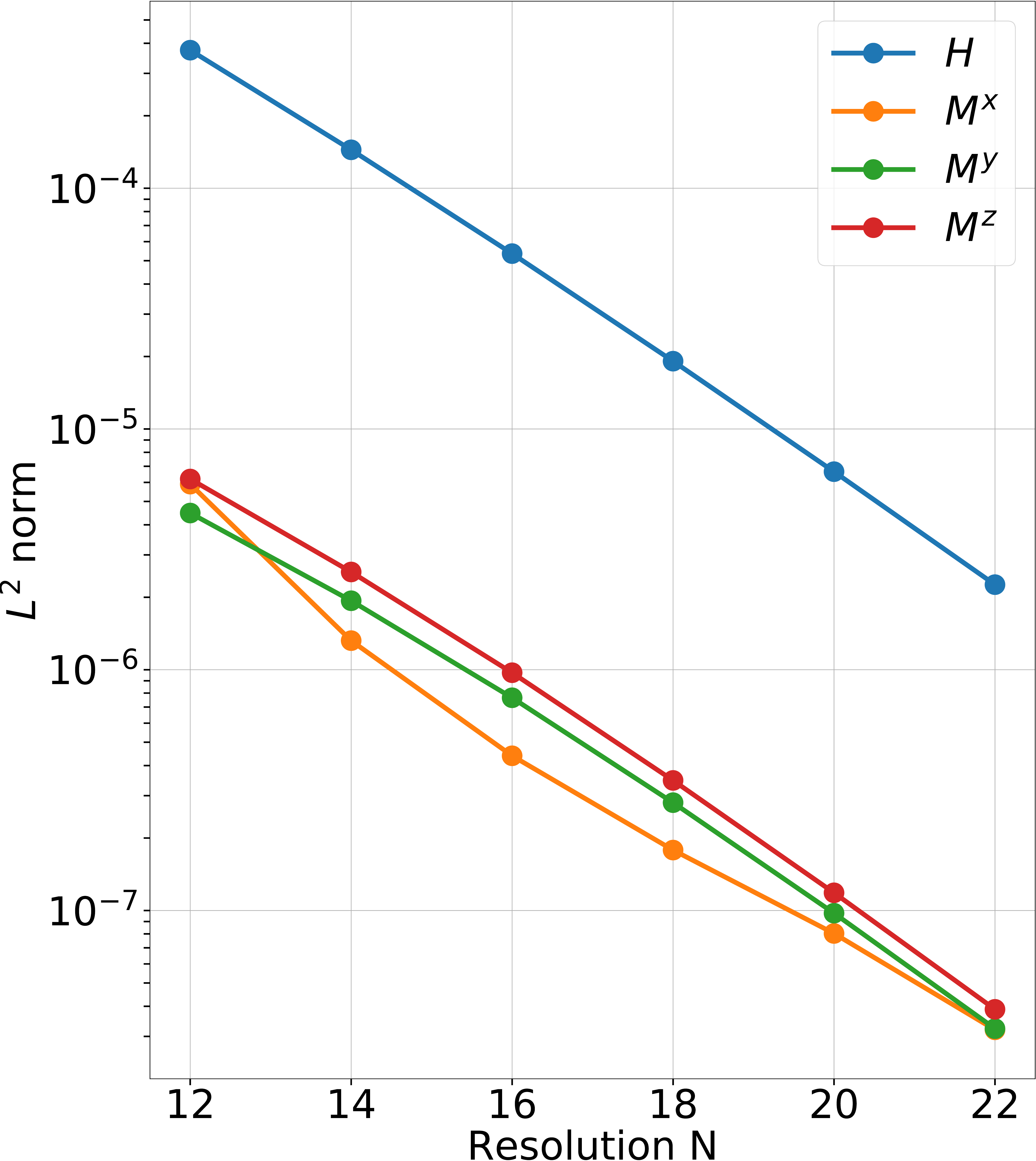}
    \caption{%
    Shown is the spectral convergence of the Hamiltonian and 
    Momentum constraints for the {\tt SXS1} ID with respect \addedaccepted{to} 
    resolution.
    \addedaccepted{The vertical axis shows the $L^2$ norm of each constraint.
    The horizontal axis shows the number of points used in each direction
    inside each cubed sphere patch.}}
    \label{fig_constraints}
\end{figure}

\begin{figure}[t]
    \includegraphics[width=1.0\linewidth]{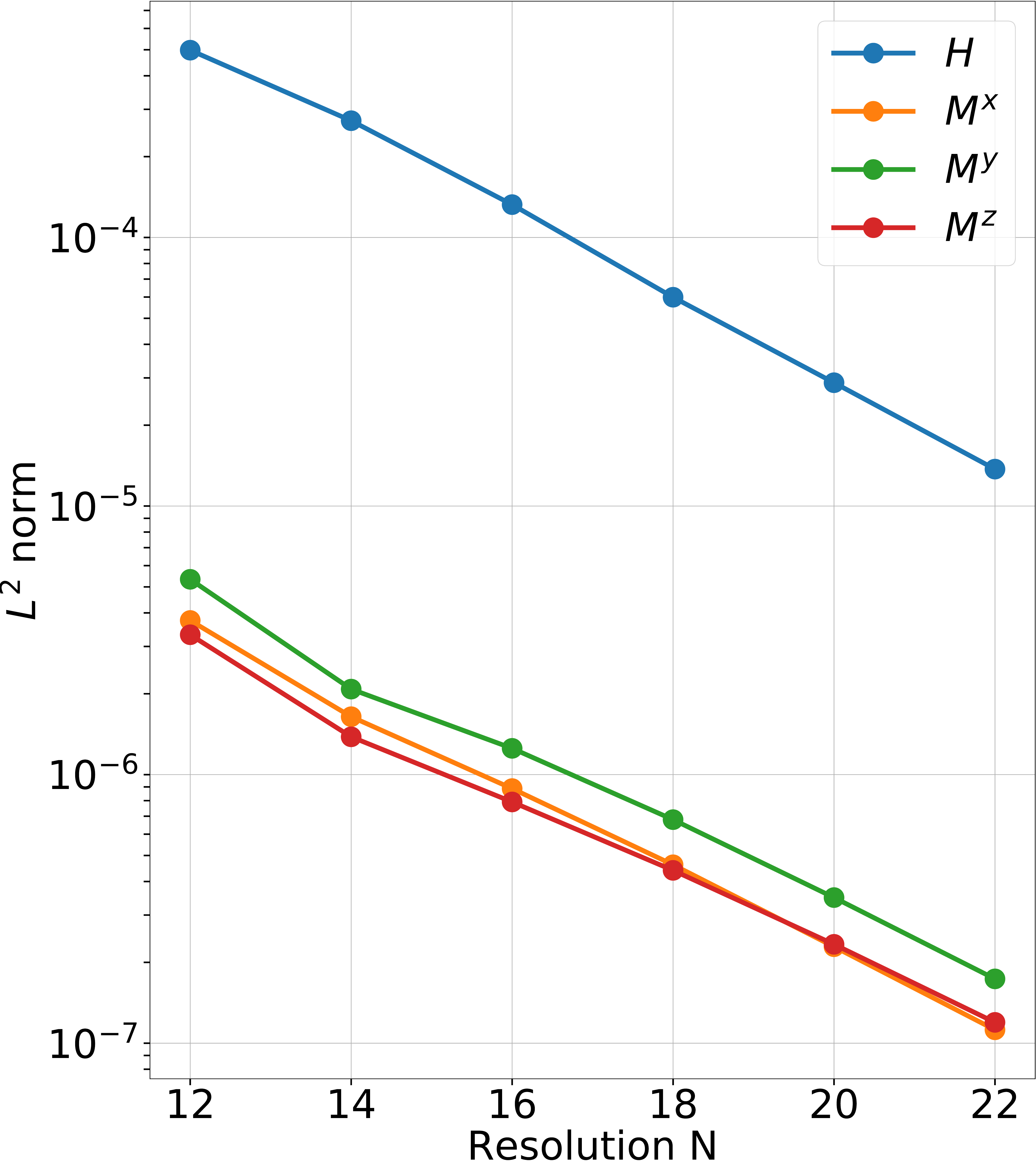}
    \caption{%
    \addedaccepted{
    Shown is the spectral convergence of the Hamiltonian and 
    Momentum constraints for the {\tt S1S2} ID with respect \addedaccepted{to} 
    resolution.
    The vertical axis shows the $L^2$ norm of each constraint.
    The horizontal axis shows the number of points in each direction in 
    a cubed spherical patch.}
    }
    \label{fig_constraints2}
\end{figure}

The first test is the verification of exponential convergence expected
for a spectral code.
We measure the violation of Hamiltonian 
and momentum constraints~\cite{Tichy_2016, gourgoulhon_thebook} using:
\begin{align}
        H := R - K_{ij}K^{ij}+K^2 - 16 \pi E
        & = 0,
        \label{eq_ham_constraint}
        \\
        M^i := D_j (K^{ij} - \gamma^{ij} K ) - 8 \pi j^i
        & = 0.
        \label{eq_mom_constraint}
\end{align}
\addedaccepted{We construct two sets of ID. 
One is {\tt SXS1} of Tab.~\ref{tab_runs} where both BH and NS 
have zero spin.
The other system, {\tt S1S2}, has $\vec \chi_{BH} = (-0.46,-0.46,-0.46)$ and 
$\vec \chi_{NS} = (0,0.32,0.32)$.
Figs.~\ref{fig_constraints} and \ref{fig_constraints2}, } show the $L^2$
norm of $H$ and $M^i$ of Eqs.~(\ref{eq_ham_constraint}) and
(\ref{eq_mom_constraint}) after the final iteration at each resolution.
Evidently, the constraint violations decay exponentially for both ID sets,
i.e., we find the expected spectral convergence.

%
%
%
%┌───────────────┐
%│ Sequence test │
%└───────────────┘
\subsection{\addedaccepted{Quasi-equilibrium Configurations}}
\label{sec_quasi_eq}
\begin{figure}[t]
	\includegraphics[width=\linewidth]{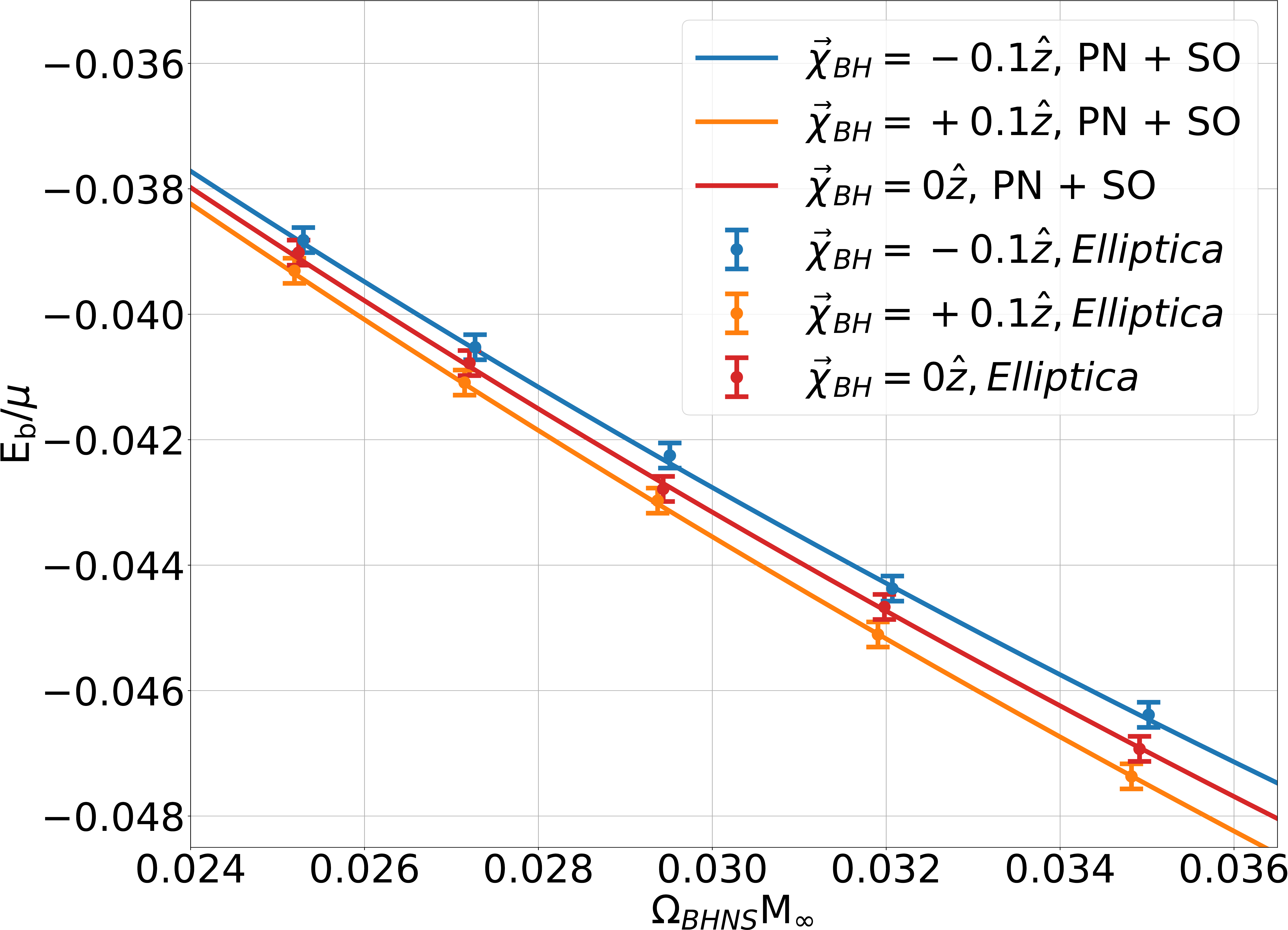}
	\caption{%
		\addedaccepted{The binding energy~($E_b$) of a BHNS system 
                versus orbital angular velocity~($\Omega_{\text{BHNS}}$),
                where the NS spin is zero and 
                $\vec \chi_{BH} = \pm 0.1\hat{z}$. 
		Here, $M_{\infty} = M_{\text{Chr}} + M_{\text{NS}}$ and 
		$\mu = \frac{ M_{\text{Chr}} M_{\text{NS}}}{M_{\infty}}$.
		The small vertical bar is the error in the numerically computed
                Eq.~(\ref{eq_adm_mass}).
		It is estimated by setting analytical ID for a Schwarzschild
		BH and then using Eq.~(\ref{eq_adm_mass}) to numerically find
                $M_{\text{ADM}}$. The difference between this
		numerical value and the known ADM mass yields
		the error.}}
	\label{fig_PN_01}
\end{figure}

\addedaccepted{We next compare our numerical ID with known
analytic approximations.
As such, we compute the binding
energy $E_b$ \footnote{
$E_b = M_{\text{ADM}} - M_{\text{Chr}}-M_{\text{NS}}$,
where $M_{\text{Chr}}$ is defined at Eq.~\ref{eq_chris_mass} and 
$M_{\text{NS}}$ is the gravitational mass of the NS in isolation.} of 
a sequence in which  
$M_{\text{NS}} = 1.31$, $M_{\text{Chr}}=5.81$, and 
$\vec \chi_{BH}=\pm 0.1\hat{z}$ are kept constant
but the orbital angular velocity ($\Omega_{\text{BHNS}}$) is varied.
We use the $3.5$ post-Newtonian~(PN) approximation from~\cite{Blanchet:2013haa}
plus the next-to-next-to-leading order correction of spin-orbit~(SO)
interaction from~\cite{Blanchet:2013haa} and compare it against our numerical
results.
As shown in Fig.~\ref{fig_PN_01} we find good agreement between 
$E_b$ computed by PN$+$SO and {\tt Elliptica}.}

\begin{figure}[t]
	\includegraphics[width=1.\linewidth]{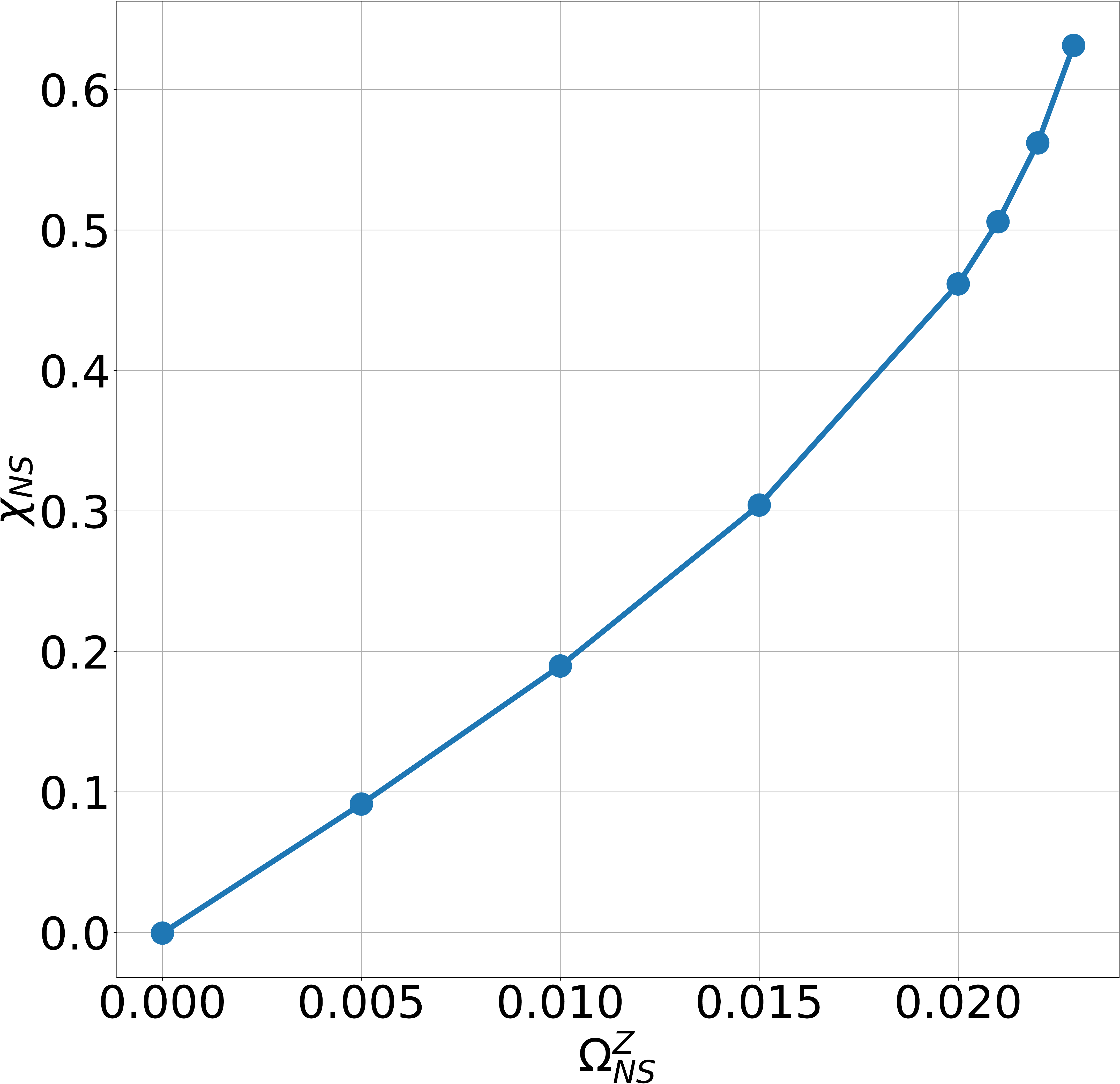}
	\caption{% 
  \addedaccepted{
	Dimensionless neutron star spin
	$\chi_{NS}$ as a function of spin-angular velocity parameter
        $\Omega^z_{\text{NS}}$. 
	Here, $\chi_{BH}=0$, $M_{\text{irr}}=5.2$, $M_{\text{B}} = 1.4$,
	and the EoS is polytropic with $K=92.12$ and $\Gamma=2$.
        The maximum $\chi_{NS}$ we can achieve is $0.63$. The breakup
        spin for an NS is approximately $0.7$~\cite{Lo2011}.}
  }
	\label{fig_chi_omega_K92}
\end{figure}

\addedaccepted{In order to test the NS spin limits in {\tt Elliptica}, 
we construct a sequence of BHNS ID where $\chi_{BH} = 0$, 
$M_{\text{irr}}=5.2$ and $M_{\text{B}} = 1.4$ are kept constant 
but $\Omega^z_{\text{NS}}$ is increased until the maximum achievable value 
($0.0229$) for this EoS is reached. This maximum $\Omega^z_{\text{NS}}$
is given by the value from which onwards
{\tt Elliptica}'s NS surface finder fails to converge
(presumably because $\Omega^z_{\text{NS}}$ then is too close to the breakup
spin of the NS).
We use Eq.~(\ref{eq_JRP}) to calculate the NS spin $\chi_{NS}$ for each
$\Omega^z_{\text{NS}}$.
As shown in Fig.~\ref{fig_chi_omega_K92}, for low values of 
$\Omega^z_{\text{NS}}$ the spin $\chi_{NS}$ is a linear function of 
$\Omega^z_{\text{NS}}$, but for higher $\Omega^z_{\text{NS}}$ the spin
rises more and more quickly.}
\subsection{Evolution Tests}
\label{sec_evolution_tests}
\begin{figure}[t]
	\centering
	\includegraphics[width=1.0\linewidth]{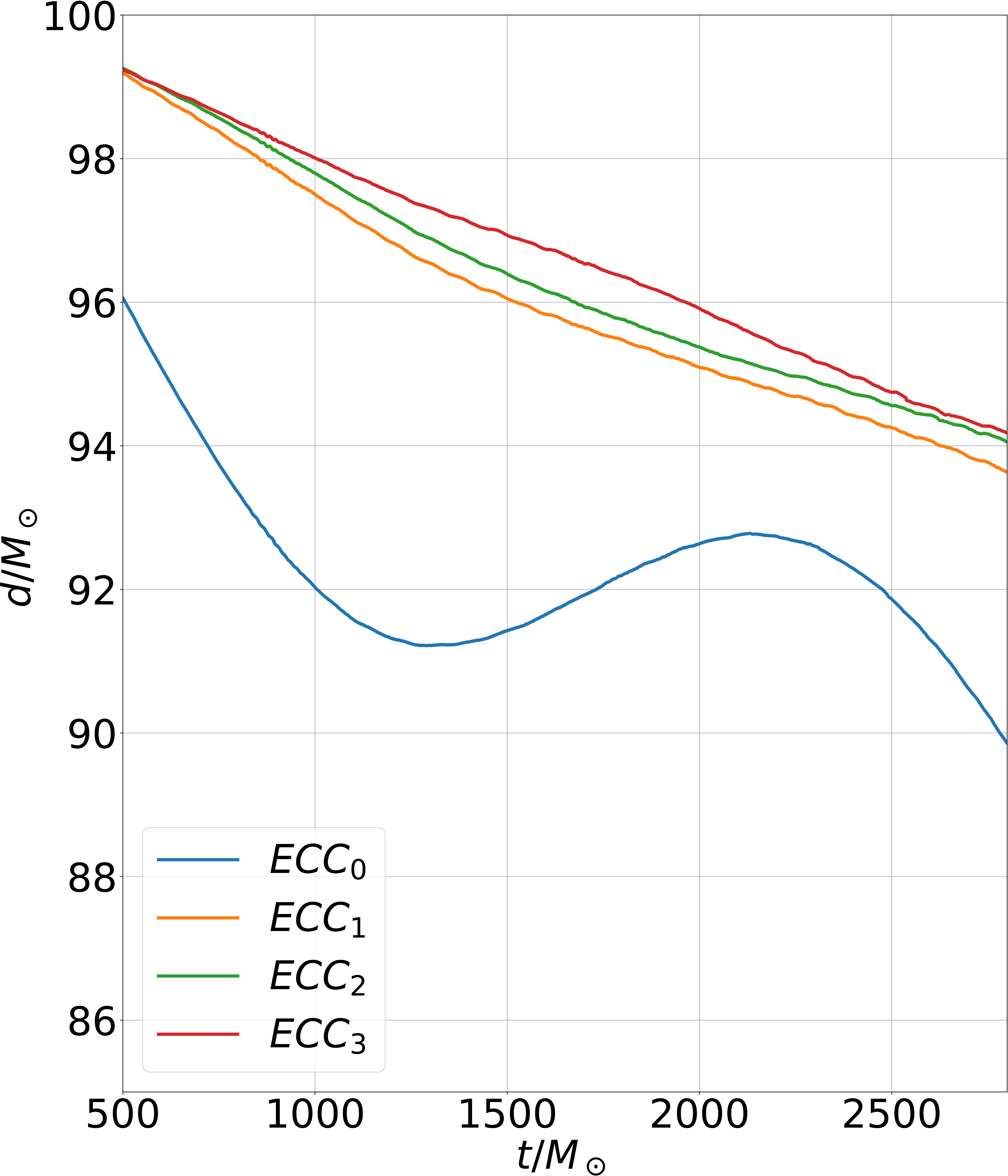}
	\caption{%
  \addedaccepted{
	The proper distance~(d) between BH and NS versus time~(t) of the {\tt SXS1}
        system for different $\Omega_{\text{BHNS}}$. After evolving for
        about $3$ orbits the eccentricity reduction algorithm 
        in~\cite{Tichy_2019} was used to adjust 
	$\Omega_{\text{BHNS}}$ and $v_r$.
        The eccentricity of each curve is as follows:
	$ECC_0 = 3.1 \times 10^{-2}$, $ECC_1 = 3.8 \times 10^{-3}$, 
        $ECC_2 = 3.4 \times 10^{-3}$, and $ECC_3 = 9.0 \times 10^{-4}$. 
        For the reduction we exclude the first $500 M_\odot$ so that 
        the required fits are not affected by the initial gauge adjustments 
	(as mentioned in \ref{sec_evolution_tests}).}
  }
	\label{fig_ecc_red}
\end{figure}

To further test the IDs, we evolve them using the {\tt BAM}
code~\cite{Chaurasia2021}.
Since for the construction of the ID we use an excision method, 
but want to use the moving puncture method for the evolution, 
we have to fill the inside of the BH with smooth data~\cite{Brown_2007,Faber_2007,Brown2008,Reifenberger_2012}.
Here, smooth means that the fields have to be at least $C^2$ 
across the apparent horizon.
\addedaccepted{
In order to have low eccentricity in the SXS1 (see Tab.~\ref{tab_runs}) 
inspiral, we perform three eccentricity-reduction steps 
as in~\cite{Tichy_2019} to obtain the target eccentricity
$\lesssim 10^{-3}$. Fig.~\ref{fig_ecc_red} shows the results of this 
eccentricity reduction algorithm for each step.}

\begin{figure}[t]
	\centering
	\includegraphics[width=1.0\linewidth]{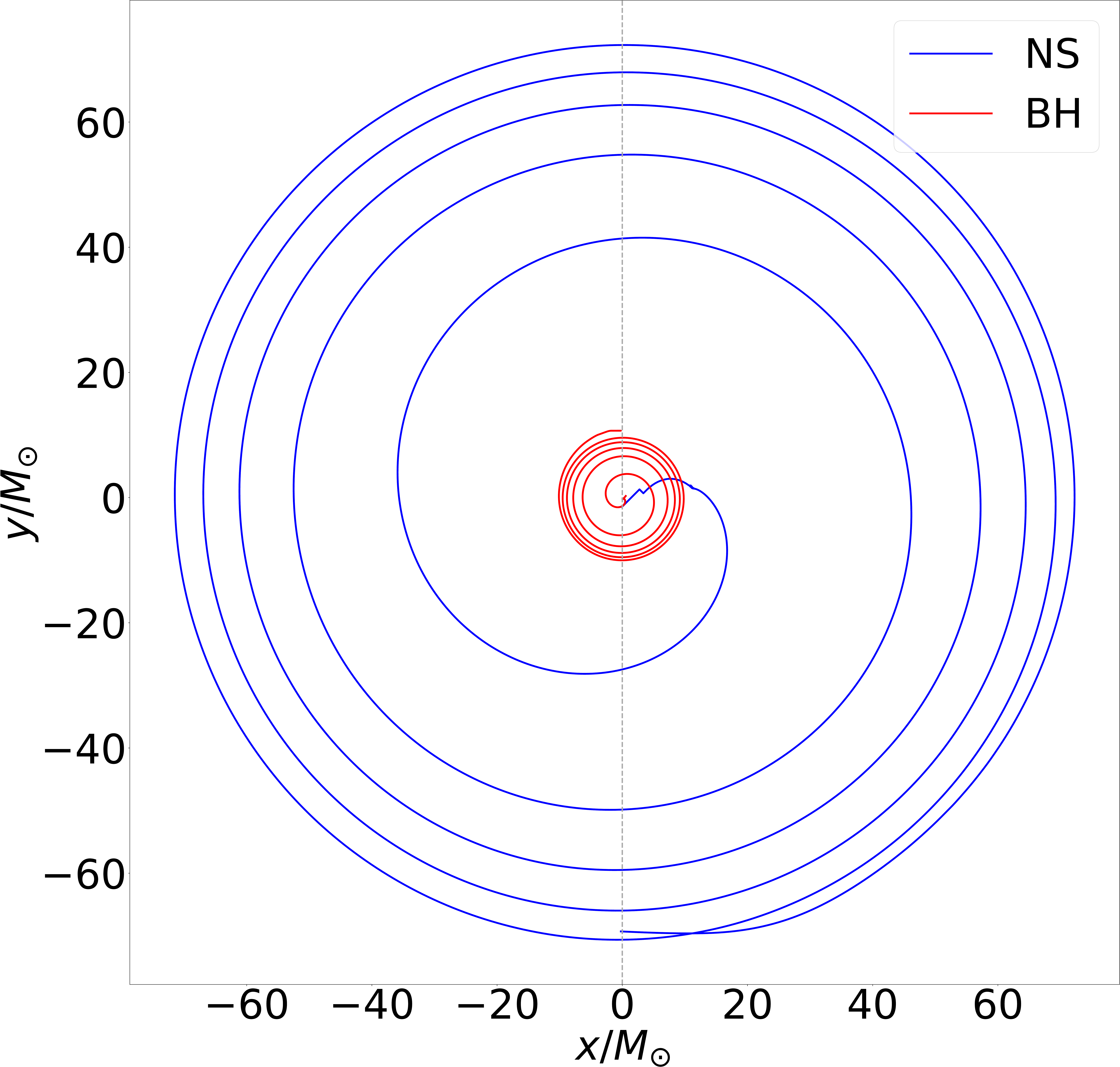}
	\caption{%
		Trajectories of BH (in red) and NS (in blue) up until 
                merger as described in \ref{sec_evolution_tests}. 
                The objects start their evolution aligned on the $x=0$ 
                axis (dashed line), as  computed from {\tt Elliptica}. 
                The tracks follow the evolution of the minimum of the 
                lapse relative to each object. \addedaccepted{The straight motion at the
                beginning of each trajectory is due to gauge adjustments.}
		}
	\label{fig_sxs1}
\end{figure}

In Fig.~\ref{fig_sxs1}, we plot
the trajectories of the BH and NS from the eccentricity reduced simulation
of {\tt SXS1}. 
As expected, they spiral in and merge on low eccentricity orbits, without
any visible drift of the center of mass.

The trajectories show an initially straight motion for both objects, 
which is related to adjustments of the gauges.
To keep the plot clean, we only show the trajectories of the BH and NS's
centers, and not the extent of the two objects.
We track the location of the NS by the minimum in the lapse. The 
small wiggles and/or jumps around merger time are mostly due to 
the minimum of the lapse no longer being an appropriate indicator for
the center of the NS.

\begin{figure}[t]
	\includegraphics[width=1.0\linewidth]{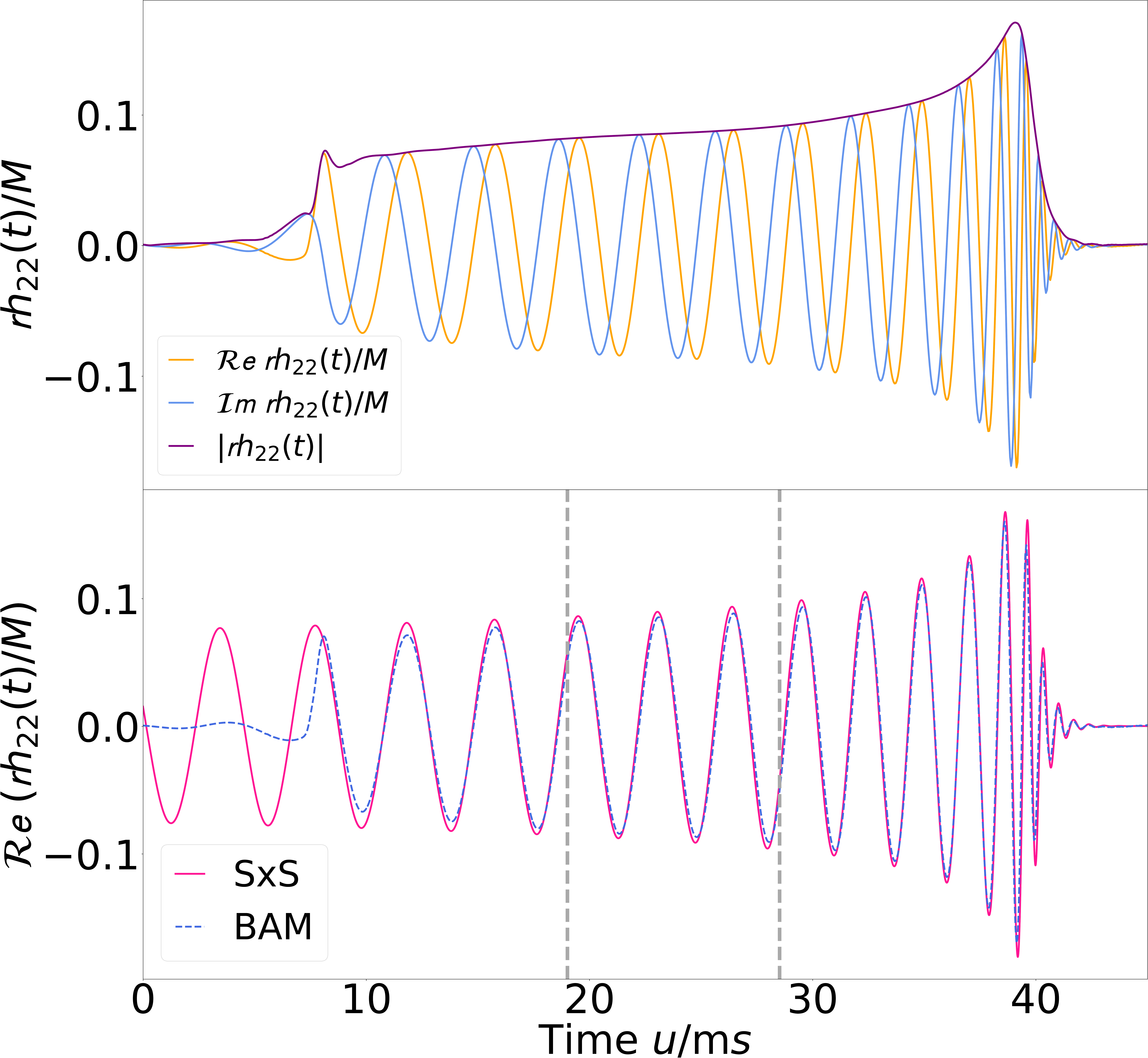}
	\caption{%
	\addedaccepted{
	\emph{Top:} Extracted gravitational wave strains~($rh_{22}$) 
        for the $l=m=2$ mode of the {\tt SXS1} system from table~\ref{tab_runs}.
        \emph{Bottom:} Comparison of $rh_{22}$ between the 
        SXS:BHNS:$0001$ system from~\cite{SxS_catalog}~(solid line)
        and the {\tt SXS1} system~(dashed line). For visual
        clarity, only the real part of both strains ($\mathcal{R}e(rh_{22})$)
        is shown in the plot.
        The alignment interval is marked with the vertical dashed line.
        Waveforms in both panels are plotted against the retarded time 
	$u$ defined at Eq.~(\ref{eq_retarded_time}) and 
        $M$ is the sum of the isolated BH and NS’s masses.} }
	\label{fig_sxs2}
\end{figure}

\addedaccepted{{\tt SXS1} has the same physical parameters as the first run in the SXS collaboration 
catalog~\cite{SxS_catalog} (SXS:BHNS:$0001$).}
In Fig.~\ref{fig_sxs2}, we depict the extracted gravitational 
wave signal emitted by this system, in the form of the dominant mode
($l=m=2$). The waveforms are plotted against the retarded time $u$ calculated as
\begin{equation}
u = t - r_* = t - r_{extr} - 2 M\ln(r_{extr}/2M - 1),
\label{eq_retarded_time}
\end{equation}
where $r_{extr}$ is the extraction radius, set to $\sim1500$ $ M_{\odot}$, and
$M$ is the sum of the isolated BH and NS's masses.
As is visible in the bottom panel, 
our result agrees with the configuration SXS:BHNS:$0001$ from the
SXS collaboration catalog~\cite{SxS_catalog}. Note that both systems share
the same  physical properties, i.e., masses, spins, and EoS, but start
from different initial separations.

\begin{figure}[t]
	\raggedleft
	\includegraphics[width=1.\linewidth]{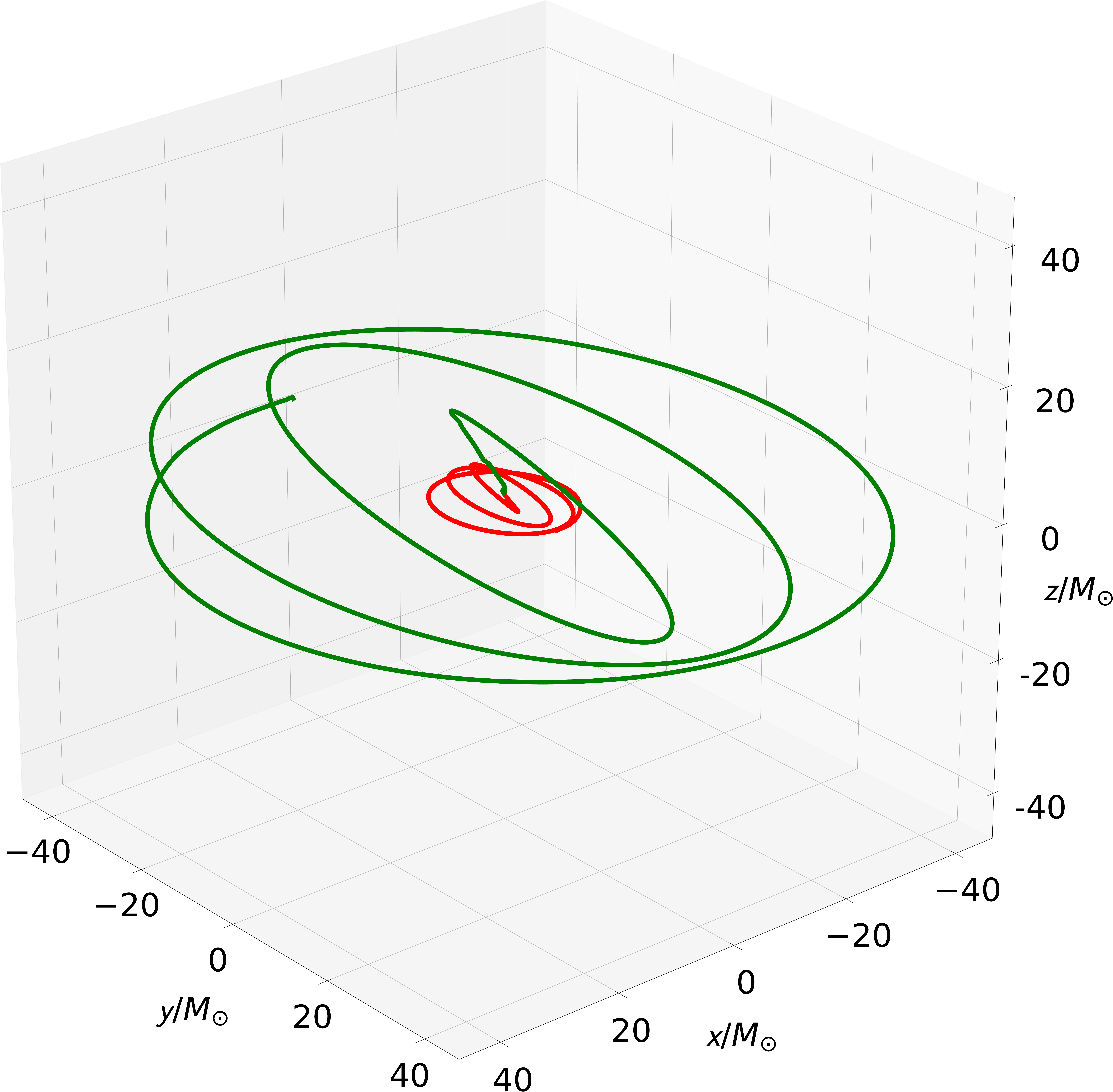}
	\caption{%
  \addedaccepted{
	 $3$-D inspiraling and precessing trajectories of 
	the NS~(green trajectory) and BH (red trajectory).
	Here, $\vec \chi_{BH} = (-0.46,-0.46,-0.46)$, 
	$\vec \chi_{NS} = (0,0.32,0.32)$, and the initial separation $s =
	56$.}
  }
	\label{fig_3D_traj}
\end{figure}

\addedaccepted{As a second example, we evolve the {\tt S1S2} BHNS system~%
(already described in Sec.~\ref{sec_spectral_convergence_test})
that has $\vec \chi_{BH} = (-0.46,-0.46,-0.46)$ and 
$\vec \chi_{NS} = (0,0.32,0.32)$.
This system presents a setup with misaligned spins (with respect 
to the initial orbital angular momentum), which leads to precession.
Fig.~\ref{fig_3D_traj} depicts the $3$-dimensional trajectories 
of the objects. The initial coordinate distance between BH and NS
is $s = 56$. This leads to $3$ orbits before the merger.
Further analysis and gravitational waves
of such systems, with longer inspiral, are left for future work.
}

%
%┌─────────┐
%│ summary │
%└─────────┘
\section{Summary}
\label{sec_summary}
The construction of ID for self-gravitating 
astrophysical systems is an indispensable task for an accurate 
dynamical evolution,
and consequently for an understanding of compact binary coalescences.
For this purpose, we have developed a new code, {\tt Elliptica}, which
provides infrastructure to solve the constraint equations and produce
ID of such astrophysical systems.

The current version of the code uses Chebyshev polynomials of the first kind 
to spectrally expand the fields over the computational grid.
This grid is covered by several cubic and cubed sphere coordinate patches.
To solve the constraint equations, 
which are coupled nonlinear elliptic PDEs,
a Newton-Raphson method is used in which 
the linearized equations are solved by a SCDD method and
the Jacobian is set \added{analytically.} 
Furthermore, the code supports polytropes and piecewise polytropic
equations of state for the NS. 
The NS can have a high spin in an arbitrary direction.
The BH can also have an arbitrary spin direction with a maximum
dimensionless spin magnitude $\sim 0.8$. 

For testing and proof of concept, we have constructed ID for
\addedaccepted{various cases of spinning and non-spinning BH and NS for 
BHNS systems. We have further verified that sequences of BHNS ID
with $\chi_{\text{BH}} = \pm 0.1\hat{z}$ agree well with 
analytical PN approximations.}
\addedaccepted{Moreover, we have constructed ID for BHNS system {\tt SXS1}
that has the same physical parameters as SXS:BHNS:$0001$ from the SXS
catalog~\cite{SxS_catalog}, as well as BHNS system {\tt S1S2} where
both BH an NS have generic spins.
For both cases we have confirmed the expected spectral convergence
of their Hamiltonian and momentum constraints.}
We have also evolved both {\tt SXS1} and {\tt S1S2} and verified that their 
orbits show the expected inspiral behavior. Furthermore, we have confirmed
that the emitted gravitational waves of the {\tt SXS1} system agree well
with results from prior studies.

In the future, we plan to achieve maximal BH spin angular momentum
by changing the conformal metric in the vicinity of the BH 
like in~\cite{Tacik2016} or use a puncture method as in~\cite{Ruchlin_2017}.

%┌─────────────────┐
%│ acknowledgments │
%└─────────────────┘

\begin{acknowledgments}
\addedaccepted{
This work was supported in part by NSF Grants
PHY-1707227, PHY-2011729, as well as DFG Grant BR 2176/5-1, and
the Coordena\c{c}\~ao de Aperfei\c{c}oamento de Pessoal 
de N\'ivel Superior- Brasil (CAPES)-Process number: 88887.571346/2020-00.
Additionally, ``Gravitational Radiation and Electromagnetic
Astrophysical Transients (GREAT)'' funded by the Swedish
Research council (VR) under Grant No. Dnr. 2016-06012 supported this project.
We also acknowledge usage of computer time on the HPC cluster KOKO
at Florida Atlantic University, 
on Lise/Emmy of the North German Supercomputing
Alliance (HLRN) [project bbp00049], on HAWK at the
High-Performance Computing Center Stuttgart (HLRS)
[project GWanalysis 44189], and on SuperMUC NG of the
Leibniz Supercomputing Centre (LRZ) [project pn29ba].
}

\end{acknowledgments}

%┌──────────┐
%│ appendix │
%└──────────┘
%\appendix

%\section{Jacobian?}
%┌──────────────┐
%│ bibliography │
%└──────────────┘
%\bibliographystyle{plain}
\bibliography{../references}

\end{document}